\documentclass[reprint,prd,twocolumn,superscriptaddress,showpacs,floatfix]{revtex4}
\usepackage{graphicx}
\usepackage{amssymb}
\usepackage{epsfig}
\usepackage{multirow}
\newcommand{\ta}[1]{#1\hspace{-.75em}/\hspace{.20em}} 

\begin{document}


\title{Modeling interactions of photons with pseudoscalar and vector mesons. }
\thanks{Work 
supported in part by
the Polish National Science Centre, grant number DEC-2012/07/B/ST2/03867
and
German Research Foundation DFG under
Contract No. Collaborative Research Center CRC-1044.}

\author{Henryk Czy\.z}
\affiliation{Institute of Physics, University of Silesia,
PL-41500 Chorz\'ow, Poland and Helmholtz-Institut, 55128 Mainz, Germany }
\author{Patrycja Kisza}
\affiliation{Institute of Physics, University of Silesia,
PL-41500 Chorz\'ow, Poland.}
\author{Szymon Tracz}
\affiliation{Institute of Physics, University of Silesia,
PL-41500 Chorz\'ow, Poland.}


\date{\today}

\begin{abstract}
  We model interaction of photons, pseudoscalars and vector mesons within the resonance
  chiral symmetric theory with the SU(3) breaking. The couplings of the model are fitted 
  to the experimental data. Within the developed model we predict the light-by-light
  contributions to the muon anomalous magnetic moment $a_{\mu}^P = (82.8 \pm 3.4)\times 10^{-11}$.
  The error covers also the model dependence within the class of models considered in this paper.
  The model was implemented into the Monte Carlo event generator Ekhara to simulate the
  reactions $e^+e^-\to e^+e^- P$, $P=\pi^0,\eta,\eta'$ and into the Monte Carlo event generator
  Phokhara to simulate the reactions $e^+e^-\to P\gamma(\gamma)$.

\end{abstract}


\maketitle

\section{Introduction}

  During the last years many very accurate experimental data,
  which contain information about photon-hadron interactions, emerged.
  In the same time one can observe a significant contribution from the theory
  community to improve the quality of the models used to describe the experimental data.
  Thus the quest for precision in hadron-photon interactions \cite{Actis:2010gg} is well under way.
  Two main reasons for this effort, besides the pure interest in knowing better the microscopic 
  world, are: the discrepancy at the level of almost 4 $\sigma$
  between the measured \cite{Bennett:2006fi} and the calculated 
 \cite{Hagiwara:2006jt,Davier:2010nc,Benayoun:2015gxa,Hagiwara:2017zod,Jegerlehner:2017gek,Davier:2017zfy}
  values of anomalous magnetic moment of the muon ($a_\mu$) and the accuracy of the electromagnetic running coupling
  constant calculated at $M_Z$ \cite{Hagiwara:2006jt}, 
   which is a limiting factor in future tests of the Standard Model.
  In both cases the hadronic contributions are the source of the uncertainties as electroweak corrections
    are well under control. 

   In this paper we extend the validity of the model developed in \cite{Czyz:2012nq} to be able 
   not only to model correctly the  $\gamma^*-\gamma^*-P$ form factors in the space-like region,
  which are necessary to calculate the light-by-light contributions to the $a_\mu$ 
   \cite{Jegerlehner:2009ry,Jegerlehner:2017gek}, but also to describe correctly all the experimental
   data which can be predicted from the Lagrangians $\mathcal{L}_{\gamma \gamma \mathcal{P}}$,
    $\mathcal{L}_{\gamma V}$, $\mathcal{L}_{ V\gamma P}$ and  $\mathcal{L}_{VVP}$. 
   A similar research program  of a global fit 
    was  carried within the Hidden Local Symmetry (HLS) effective Lagrangian
   \cite{Benayoun:2011mm,Benayoun:2012wc,Benayoun:2015gxa} with many statistical test carried,
    yet concentrating on the modeling
   of the processes needed for the calculations of the leading order hadronic vacuum polarization
   contributions to $a_\mu$. 
    We plan to extend our analysis to cover also the $e^+e^- \to \pi^+\pi^-, K^+K^-, 
   K^0\bar K^0 \ {\rm and}, \pi^+\pi^-\pi^0$ in a future publication. This way it will be possible
  to study model dependence of the obtained results, comparing the HLS and the resonance chiral Lagrangian
   approach, which despite similarities are not identical.  
    The $\gamma^*-\gamma^*-P$ form factors, one of the outcome of this paper,
   are modeled within various frameworks 
   \cite{Feldmann:1997vc,Kroll:2002nt,Scarpettini:2003fj,Agaev:2003kb,Noguera:2011fv,Masjuan:2012wy,Stefanis:2012yw,Noguera:2012aw,Wu:2012kw,Klopot:2012hd,Czyz:2012nq,Geng:2012qg,Dumm:2013zoa,deMelo:2013zza,Li:2013xna,Escribano:2013kba,Agaev:2014wna,Escribano:2015nra,Escribano:2015yup,Zhong:2015nxa,GomezDumm:2016bxp}: phenomenology oriented, aiming for model independence
   Pad\'e approximants, chiral effective resonance theory, quark models and Nambu-Jona-Lasinio model.

   The paper is organized in the following way: In Section \ref{model} we describe the modifications
  of the model developed in \cite{Czyz:2012nq}. In Section \ref{fits} we describe the fits to 
  experimental data. In Section \ref{asymp} the 
   asymptotic behaviour and the slopes 
   of the pseudoscalar form factors
   are discussed. In Section \ref{amu} we present the evaluation, within the developed model,
  of the light-by-light contributions to the anomalous magnetic moment of the muon. In Section 
   \ref{phek} the implementations to the Monte Carlo event generators Phokhara 
   \cite{Rodrigo:2001kf,Czyz:2016xvc} and 
    Ekhara \cite{Czyz:2010sp,Czyz:2006dm}
   are presented. We shortly summarize the results in Section \ref{conc}.

\section{The model} \label{model}

  As said already in the Introduction, one of the aims
    of this paper was to extend the model used in \cite{Czyz:2012nq}
  for modeling of the $\gamma^*-\gamma^*-P$ form factors in the space-like
   region to be able to cover also the time-like region, adding to the list
  of modeled entities also other physical
  observables (see Section \ref{fits}). In \cite{Czyz:2012nq}
   the SU(3) symmetry was assumed for the couplings in the relevant 
   Lagrangians. However from the experimental data,
   which are modeled by the form factors in the time-like region, 
   it is evident that this symmetry 
  is broken (see the discussion in the next section). 
    The strategy to model all the space-like and the time-like data
   was to extend the model from \cite{Czyz:2012nq} in the minimal
   possible way to describe the whole set of the experimental data.
   In \cite{Czyz:2012nq} it was checked that the space-like data can 
   be modeled using only two vector-meson octets. When extending the model
   to the time-like region as well, one has to use at least three octets.
   This was adopted within this paper. The $\eta-\eta'$ mixing scheme,
   which was taken in \cite{Czyz:2012nq} from\cite{Feldmann:1999uf,Feldmann:1998vh}, is kept unchanged. However, as there are new data available, we have fitted
 the mixing parameters to the experimental observables predicted 
  from the Lagrangians described below.

The Wess-Zumino-Witten Lagrangian \cite{Wess:1971yu,Witten:1983tw}, which describes the interaction of pseudoscalar mesons with two photons, can be written down in the terms of the physical fields as
\begin{eqnarray}
\label{l0}
\mathcal{L}_{\gamma \gamma \mathcal{P}}&=&\frac{-e^2N_c}{24\pi^2f_{\pi}}\epsilon^{\mu\nu\alpha\beta}\partial_{\mu}B_{\nu}\partial_{\alpha}B_{\beta}\Big[\pi^0+\eta(\frac{5}{3}C_q-\frac{\sqrt{2}}{3}C_s)\nonumber\\
&+&\eta^{\prime}(\frac{5}{3}C_q^{\prime}+\frac{\sqrt{2}}{3}C_s^{\prime})\Big]\, .
\end{eqnarray}

The $\gamma V$ interaction is described in terms of the following Lagrangian:
\begin{equation}
\label{l1}
\mathcal{L}_{\gamma V}=-e\sum_{i=1}^3 f_{V_i}\partial_{\mu}B_{\nu}\Big(\tilde{\rho_i}^{\mu\nu}+\frac{1}{3}F_{\omega_i}\tilde{\omega}_i^{\mu\nu}-\frac{\sqrt{2}}{3}F_{\phi_i}\tilde{\phi}_i^{\mu\nu}\Big),
\end{equation}
where $\tilde{V}_{\mu\nu}\equiv \partial_{\mu}V_{\nu}-\partial_{\nu}V_{\mu}$, $f_{V_i}$ is dimensionless coupling for the vector representation of the spin-1 fields 
in a given octet. The SU(3) symmetry of the coupling constants is broken 
  here in the first octet only, by introducing the additional constants
  $F_{\omega_1}$ and $F_{\phi_1}$. For the other octets the constants are set to
  1: $F_{\omega_i}=F_{\phi_i}=1,{\rm for}\  i=2,3$ preserving the SU(3) symmetry
   in the higher octets. 

The Lagrangians that describe vector-photon-pseudoscalar and two vector mesons interaction with pseudoscalar come from extension of the Lagrangians
 from \cite{Prades:1993ys}, which were adopted in \cite{Czyz:2012nq}. 
 In terms of the physical fields they read
\begin{widetext}
\begin{equation}
\label{l2}
\mathcal{L}_{V\gamma \pi^0}=-\sum_{i=1}^n \frac{4\sqrt{2}eh_{V_i}}{3f_{\pi}}\epsilon_{\mu\nu\alpha\beta}\partial^{\alpha}B^{\beta}\Big(\rho_i^{\mu}+3H_{\omega_i}\omega_i^{\mu}-\frac{3}{\sqrt{2}}A_i^{\pi_0}\phi_i^{\mu}\Big)\partial^{\nu}\pi^0\, ,
\end{equation}
\begin{equation}
\label{l3}
\mathcal{L}_{V\gamma \eta}=-\sum_{i=1}^n \frac{4\sqrt{2}eh_{V_i}}{3f_{\pi}}\epsilon_{\mu\nu\alpha\beta}\partial^{\alpha}B^{\beta}\Big[(3\rho_i^{\mu}+\omega_i^{\mu})C_q+2\phi_i^{\mu}C_s-(\frac{5}{\sqrt{2}}C_q-C_s)A_i^{\eta}\phi_i^{\mu}\Big]\partial^{\nu}\eta\, ,
\end{equation}
\begin{equation}
\mathcal{L}_{V\gamma \eta^{\prime}}=-\sum_{i=1}^n \frac{4\sqrt{2}eh_{V_i}}{3f_{\pi}}\epsilon_{\mu\nu\alpha\beta}\partial^{\alpha}B^{\beta}\Big[(3\rho_i^{\mu}+\omega_i^{\mu})C^{\prime}_q-2\phi_i^{\mu}C^{\prime}_s-(\frac{5}{\sqrt{2}}C^{\prime}_q+C^{\prime}_s)A_i^{\eta^{\prime}}\phi_i^{\mu}\Big]\partial^{\nu}\eta^{\prime}\, ,
\label{l4}
\end{equation}
\begin{eqnarray}
\mathcal{L}_{VV\pi^0}&=&-\sum_{i=1}^n\frac{4\sigma_{V_i}}{f_{\pi}}\epsilon_{\mu\nu\alpha\beta}\Big[\frac{1}{F_{\omega_i}}\pi^0\partial^{\mu}\omega_i^{\nu}\partial^{\alpha}\rho_i^{\beta}+\frac{3(F_{\omega_i}H_{\omega_i}-1-A^{\pi^0}_{\phi\omega, i})}{2F_{\omega_i}^2}\pi^0\partial^{\mu}\omega_i^{\nu}\partial^{\alpha}\omega_i^{\beta} \nonumber\\
&+&\frac{3(A_i^{\pi_0}-A^{\pi^0}_{\phi\omega, i}/F_{\phi_i})}{4F_{\phi_i}}\pi^0\partial^{\mu}\phi_i^{\nu}\partial^{\alpha}\phi_i^{\beta}
-\frac{3A^{\pi^0}_{\phi\omega, i}}{\sqrt{2}F_{\omega_i}F_{\phi_i}}\pi^0\partial^{\mu}\phi_i^{\nu}\partial^{\alpha}\omega_i^{\beta}\Big],
\label{l5}
\end{eqnarray}
\begin{eqnarray}
\mathcal{L}_{VV\eta}&=&-\sum_{i=1}^n\frac{4\sigma_{V_i}}{f_{\pi}}\epsilon_{\mu\nu\alpha\beta}\eta\Big[(\partial^{\mu}\rho_i^{\nu}\partial^{\alpha}\rho_i^{\beta}+\frac{1}{F_{\omega_i}}\partial^{\mu}\omega_i^{\nu}\partial^{\alpha}\omega_i^{\beta})\frac{1}{2}C_q
-\frac{9A^{\eta}_{\phi\omega_i}}{F^2_{\omega_i}}\partial^{\mu}\omega_i^{\nu}\partial^{\alpha}\omega_i^{\beta}
-\frac{1}{F_{\phi_i}}\partial^{\mu}\phi_i^{\nu}\partial^{\alpha}\phi_i^{\beta}\frac{1}{\sqrt{2}}C_s \nonumber\\
&-&\frac{9A^{\eta}_{\phi\omega, i}}{2F^2_{\phi_i}}\partial^{\mu}\phi_i^{\nu}\partial^{\alpha}\phi_i^{\beta}
+\frac{A_i^{\eta}}{6F_{\phi_i}}(\frac{15}{2}C_q-\frac{3}{\sqrt{2}}C_s)\partial^{\mu}\phi_i^{\nu}\partial^{\alpha}\phi_i^{\beta}
-\frac{9\sqrt{2}A^{\eta}_{\phi\omega,i}}{F_{\omega_i} F_{\phi_i}}\partial^{\mu}\phi_i^{\nu}\partial^{\alpha}\omega_i^{\beta}\Big],
\label{l6}
\end{eqnarray}

\begin{eqnarray}
\mathcal{L}_{VV\eta^{\prime}}&=&-\sum_{i=1}^n\frac{4\sigma_{V_i}}{f_{\pi}}\epsilon_{\mu\nu\alpha\beta}\eta^{\prime}\Big[(\partial^{\mu}\rho_i^{\nu}\partial^{\alpha}\rho_i^{\beta}+\frac{1}{F_{\omega_i}}\partial^{\mu}\omega_i^{\nu}\partial^{\alpha}\omega_i^{\beta})\frac{1}{2}C^{\prime}_q
+\frac{1}{F_{\phi_i}}\partial^{\mu}\phi_i^{\nu}\partial^{\alpha}\phi_i^{\beta}\frac{1}{\sqrt{2}}C^{\prime}_s \nonumber\\
&+&\frac{A_i^{\eta^{\prime}}}{6F_{\phi_i}}(\frac{15}{2}C^{\prime}_q+\frac{3}{\sqrt{2}}C^{\prime}_s)\partial^{\mu}\phi_i^{\nu}\partial^{\alpha}\phi_i^{\beta}\Big],
\label{l7}
\end{eqnarray}
\end{widetext}
where $n=3$, $H_{\omega_i}, F_{\phi_i}=1$ for $i=2,3$, 
$A^{P}_{\phi\omega, i}\neq0$ only for $i=1$ and $P=\pi^0,\eta$.  $C_q$, $C^{\prime}_q$, $C_s$, $C^{\prime}_s$ are given by the following formulae
\begin{equation}
C_q = \frac{f_{\pi}}{\sqrt{3}\cos{(\theta_8 - \theta_0)}}\Big(\frac{1}{f_8}\cos{\theta_0} -\frac{1}{f_0} \sqrt{2}\sin{\theta_8}\Big)\, ,
\label{cq}
\end{equation}
\begin{equation}
C_s = \frac{f_{\pi}}{\sqrt{3}\cos{(\theta_8 - \theta_0)}}\Big(\frac{1}{f_8}\sqrt{2}\cos{\theta_0} +\frac{1}{f_0}\sin{\theta_8}\Big)\, ,
\label{cs}
\end{equation}

\begin{equation}
C^{\prime}_q = \frac{f_{\pi}}{\sqrt{3}\cos{(\theta_8 - \theta_0)}}\Big(\frac{1}{f_0}\sqrt{2}\cos{\theta_8} +\frac{1}{f_8}\sin{\theta_0}\Big)\, ,
\label{cqp}
\end{equation}

\begin{equation}
C^{\prime}_s = \frac{f_{\pi}}{\sqrt{3}\cos{(\theta_8 - \theta_0)}}\Big(\frac{1}{f_0}\cos{\theta_8} -\frac{1}{f_8}\sqrt{2}\sin{\theta_0}\Big) .
\label{csp}
\end{equation}

 The model from \cite{Czyz:2012nq} is recovered setting $n=2$,
$H_{\omega_i}=F_{\phi_i}=1$,  $A_i^{P}=0$ and $A^{P}_{\phi\omega, i}=0$. The couplings in the
Lagrangians $\mathcal{L}_{VVP}$ are chosen to fulfil the asymptotic behaviour of the
  $P-\gamma^*-\gamma^*$ form factors. It is  discussed later in this Section.

From the Lagrangians, Eqs.(\ref{l0}-\ref{l7}), one derives the $P-\gamma^*-\gamma^*$
amplitude
\begin{equation}
\mathcal{M}[P\to \gamma^*(q_1)\, \gamma^*(q_2)  ]
= 
e^2 \epsilon_{\mu\nu\alpha\beta} q_{1}^{\mu} q_{2}^{\alpha} 
F_{\gamma^*\gamma^* P}(t_1, t_2)
.\label{amp1}
\end{equation}
\noindent
The form factors $F_{\gamma^*\gamma^* P}(t_1, t_2)$ read
\begin{widetext}
\begin{eqnarray}
F_{\gamma^* \gamma^*\pi^0}(t_1,t_2)&=&-\frac{N_c}{12\pi^2f_{\pi}}+\sum_{i=1}^n\frac{4\sqrt{2}h_{V_i}f_{V_i}}{3f_{\pi}}t_1\Big(D_{\rho_i}(t_1)+F_{\omega_i}H_{\omega_i}D_{\omega_i}(t_1)+A_i^{\pi_0}F_{\phi_i}D_{\phi_i}(t_1)\Big)\nonumber\\
&+& \sum_{i=1}^n\frac{4\sqrt{2}h_{V_i}f_{V_i}}{3f_{\pi}}t_2\Big(D_{\rho_i}(t_2)+F_{\omega_i}H_{\omega_i}D_{\omega_i}(t_2)+A_i^{\pi_0}F_{\phi_i}D_{\phi_i}(t_2)\Big)\nonumber\\
&-& \sum_{i=1}^n\frac{4\sigma_{V_i}f_{V_i}^2}{3f_{\pi}}t_1t_2\Bigg(D_{\rho_i}(t_2)D_{\omega_i}(t_1)+D_{\rho_i}(t_1)D_{\omega_i}(t_2)+
(A_i^{\pi_0}F_{\phi_i}-A^{\pi^0}_{\phi\omega,i})D_{\phi_i}(t_1)D_{\phi_i}(t_2)\nonumber\\
&+&\Big(F_{\omega_i}H_{\omega_i}-1-A^{\pi^0}_{\phi\omega,i}\Big)D_{\omega_i}(t_1)D_{\omega_i}(t_2)
+A^{\pi^0}_{\phi\omega,i}\Big(D_{\phi_i}(t_1)D_{\omega_i}(t_2)+D_{\phi_i}(t_1)D_{\omega_i}(t_2)\Big) \Bigg)\, ,\label{ffpi0}
\end{eqnarray}

\begin{eqnarray}
F_{\gamma^* \gamma^*\eta}(t_1,t_2)&=&-\frac{N_c}{12\pi^2f_{\pi}}\Big(\frac{5}{3}C_q-\frac{\sqrt{2}}{3}C_s\Big)\nonumber\\
&&\kern-150pt+\sum_{i=1}^n\frac{4\sqrt{2}h_{V_i}f_{V_i}}{3f_{\pi}}t_1\Bigg(\Big(3C_qD_{\rho_i}(t_1)+\frac{1}{3}F_{\omega_i}C_qD_{\omega_i}(t_1)-\frac{2\sqrt{2}}{3}C_sF_{\phi_i}D_{\phi_i}(t_1)\Big)
+\Big(\frac{5}{3}C_q-\frac{\sqrt{2}}{3}C_s\Big)A_i^{\eta}F_{\phi_i}D_{\phi_i}(t_1)\Bigg)\nonumber\\
&&\kern-150pt+\sum_{i=1}^n\frac{4\sqrt{2}h_{V_i}f_{V_i}}{3f_{\pi}}t_2\Bigg(\Big(3C_qD_{\rho_i}(t_2)+\frac{1}{3}C_qF_{\omega_i}D_{\omega_i}(t_2)-\frac{2\sqrt{2}}{3}C_sF_{\phi_i}D_{\phi_i}(t_2)\Big)
+\Big(\frac{5}{3}C_q-\frac{\sqrt{2}}{3}C_s\Big)A_i^{\eta}F_{\phi_i}D_{\phi_i}(t_2)\Bigg)\nonumber\\
&&\kern-150pt-\sum_{i=1}^n\frac{8\sigma_{V_i}f_{V_i}^2}{f_{\pi}}t_1t_2\Bigg[\Big(\frac{1}{2}C_qD_{\rho_i}(t_1)D_{\rho_i}(t_2)+\frac{1}{18}F_{\omega_i}C_qD_{\omega_i}(t_1)D_{\omega_i}(t_2)
-A^{\eta}_{\phi\omega,i}D_{\omega_i}(t_1)D_{\omega_i}(t_2)
-\frac{\sqrt{2}}{9}C_sF_{\phi_i}D_{\phi_i}(t_1)D_{\phi_i}(t_2) \Big)\nonumber\\
&&\kern-150pt+\frac{A_i^{\eta}F_{\phi_i}}{6}\Big(\frac{5}{3}C_q-\frac{\sqrt{2}}{3}C_s\Big)D_{\phi_i}(t_1)D_{\phi_i}(t_2)
-A^{\eta}_{\phi\omega,i}D_{\phi_i}(t_1)D_{\phi_i}(t_2)
+A^{\eta}_{\phi\omega,i}\Big(D_{\phi_i}(t_1)D_{\omega_i}(t_2)+D_{\phi_i}(t_1)D_{\omega_i}(t_2)\Big)\Bigg]\, ,\label{ffeta}
\end{eqnarray}

and
\begin{eqnarray}
F_{\gamma^* \gamma^*\eta^{\prime}}(t_1,t_2)&=&-\frac{N_c}{12\pi^2f_{\pi}}\Big(\frac{5}{3}C^{\prime}_q+\frac{\sqrt{2}}{3}C^{\prime}_s\Big)\nonumber\\
&&\kern-150pt+\sum_{i=1}^n\frac{4\sqrt{2}h_{V_i}f_{V_i}}{3f_{\pi}}t_1\Bigg(\Big(3C^{\prime}_qD_{\rho_i}(t_1)+\frac{1}{3}F_{\omega_i}C^{\prime}_qD_{\omega_i}(t_1)+\frac{2\sqrt{2}}{3}C^{\prime}_sF_{\phi_i}D_{\phi_i}(t_1)\Big)
+\Big(\frac{5}{3}C^{\prime}_q+\frac{\sqrt{2}}{3}C^{\prime}_s\Big)A_i^{\eta^{\prime}}F_{\phi_i}D_{\phi_i}(t_1)\Bigg)\nonumber\\
&&\kern-150pt+\sum_{i=1}^n\frac{4\sqrt{2}h_{V_i}f_{V_i}}{3f_{\pi}}t_2\Bigg(\Big(3C^{\prime}_qD_{\rho_i}(t_2)+\frac{1}{3}F_{\omega_i}C^{\prime}_qD_{\omega_i}(t_2)+\frac{2\sqrt{2}}{3}C^{\prime}_sF_{\phi_i}D_{\phi_i}(t_2)\Big)
+\Big(\frac{5}{3}C^{\prime}_q+\frac{\sqrt{2}}{3}C^{\prime}_s\Big)A_i^{\eta^{\prime}}F_{\phi_i}D_{\phi_i}(t_2)\Bigg)\nonumber\\
&&\kern-150pt-\sum_{i=1}^n\frac{8\sigma_{V_i}f_{V_i}^2}{f_{\pi}}t_1t_2\Bigg[\Big(\frac{1}{2}C^{\prime}_qD_{\rho_i}(t_1)D_{\rho_i}(t_2)+\frac{1}{18}F_{\omega_i}C^{\prime}_qD_{\omega_i}(t_1)D_{\omega_i}(t_2)
+\frac{\sqrt{2}}{9}C^{\prime}_sF_{\phi_i}D_{\phi_i}(t_1)D_{\phi_i}(t_2) \Big)\nonumber\\
&+& \frac{A_i^{\eta^{\prime}}F_{\phi_i}}{6}\Big(\frac{5}{3}C^{\prime}_q+\frac{\sqrt{2}}{3}C^{\prime}_s\Big)D_{\phi_i}(t_1)D_{\phi_i}(t_2)
\Bigg]\, , \label{ffetap}
\end{eqnarray}
\end{widetext}
\noindent
where the vector meson propagators $D_{V_i}(Q^2)$ in the space-like region are defined by:
\begin{equation}
D_{V_i}(Q^2)=[Q^2-M_{V_i}^2]^{-1} \, .
\end{equation}
In the  time-like region we use the propagators $D_{V_i}(Q^2)$ in the following form:
\begin{equation}
D_{V_i}(Q^2)=[Q^2-M_{V_i}^2+i\sqrt{Q^2}\Gamma_{V_i}]^{-1} \, .
\label{bwt}
\end{equation}
All masses and widths are fixed to their PDG \cite{Olive:2016xmw} values.
We require that the form factors $F_{\gamma^* \gamma^*P}(t_1,t_2)$ vanish, for any value of $t_2$ ($t_1$), when the photon virtuality $t_1$ ($t_2$) goes to infinity .
 This constraint leads to the following relations between the couplings:
\begin{equation}
-\frac{N_c}{4\pi^2}+4\sqrt{2}\sum_{i=1}^nh_{V_i}f_{V_i}(1+F_{\omega_i}H_{\omega_i}+A_i^{\pi^0}F_{\phi_i})=0,\label{as1}
\end{equation}
\begin{equation}
\sqrt{2} h_{V_i}f_{V_i}-\sigma_{V_i}f_{V_i}^2=0,\hspace{2.cm} i=1,..,n\label{as2}
\end{equation}

\begin{widetext}
\begin{equation}
\kern-50pt-\frac{N_c}{4\pi^2}(\frac{5}{3}C_q-\frac{\sqrt{2}}{3}C_s)+4\sqrt{2}\sum_{i=1}^nh_{V_i}f_{V_i}\Big[(3C_q+\frac{1}{3}F_{\omega_i}C_q-\frac{2\sqrt{2}}{3}C_sF_{\phi_i})
+(\frac{5}{3}C_q-\frac{\sqrt{2}}{3}C_s)A_i^{\eta}F_{\phi_i}\Big]=0 \, ,
\label{as3}
\end{equation}

and
\begin{equation}
\kern-50pt-\frac{N_c}{4\pi^2}(\frac{5}{3}C^{\prime}_q+\frac{\sqrt{2}}{3}C^{\prime}_s)+4\sqrt{2}\sum_{i=1}^nh_{V_i}f_{V_i}\Big[(3C^{\prime}_q+\frac{1}{3}F_{\omega_i}C^{\prime}_q+\frac{2\sqrt{2}}{3}C^{\prime}_sF_{\phi_i})
+(\frac{5}{3}C^{\prime}_q+\frac{\sqrt{2}}{3}C^{\prime}_s)A_i^{\eta^{\prime}}F_{\phi_i}\Big]=0\, .
\label{as4}
\end{equation}
\end{widetext}
These relations allow us to determine  six of the model parameters. 
  We have chosen $\sigma_{V_i}f_{V_i}^2 \ (i=1,2,3)$,  $h_{V_3}f_{V_3}$, $A^{\eta}_2$ and $A^{\eta^{\prime}}_2$ to be determined by using the asymptotic relations Eqs.(\ref{as2}), Eq.(\ref{as1})
 Eq.(\ref{as3}) and Eq.(\ref{as4}) correspondingly. 
 Remaining parameters have been fitted to experimental data.
From the Lagrangians Eqs.(\ref{l0}-\ref{l7}) one can derive also 
 the $V-P- \gamma^*$
amplitudes
\begin{equation}
\mathcal{M}[V(P)\to P(V) (q_1)\, \gamma^*(q_2)  ]
= 
e \epsilon_{\mu\nu\beta\alpha} q_{1}^{\nu} q_{2}^{\alpha} 
F_{V P \gamma^* }(t_1) \nonumber \\
,\label{amp1a}
\end{equation}
\noindent
where $t_1=q_2^2$.

The form factors, given here only for the specific channels used in the fits, 
 have the following form
\begin{eqnarray}
F_{\rho\pi^0\gamma^*}(t_1)=\frac{4\sqrt{2}h_{V_1}}{3f_{\pi}}\Big\{1-t_1D_{\omega_1}(t_1)\Big\} \, ,
\label{rhopigam}
\end{eqnarray}
\begin{widetext}
\begin{eqnarray}
F_{\omega\pi^0\gamma^*}(t_1)=\frac{4\sqrt{2}h_{V_1}}{f_{\pi}}
\Big\{H_{\omega_1}-\frac{t_1}{F_{\omega_1}}\Big[{D_{\rho_1}(t_1)}
+ {\big(H_{\omega_1}F_{\omega_1}-1-A^{\pi^0}_{\phi\omega,1}\big)} D_{\omega_1}(t_1) 
+{A^{\pi^0}_{\phi\omega,1}}D_{\phi_1}(t_1)
\Big]\Big\}\, ,
\label{omegapigam}
\end{eqnarray}
\begin{eqnarray}
 F_{\phi\pi^0\gamma^*}(t_1)= \frac{-4h_{V_1}}{f_{\pi}} 
\Big\{ A^{\pi^0}_{1}-\frac{A^{\pi^0}_{\phi\omega,1}}{F_{\phi_1}}t_1D_{\omega_1}(t_1)-{\big(A^{\pi^0}_{1}-\frac{A^{\pi^0}_{\phi\omega,1}}{F_{\phi_1}}\big)}t_1D_{\phi_1}(t_1)
\Big\}\, ,
\label{phipigam}
\end{eqnarray}

\begin{eqnarray}
 F_{\rho\eta\gamma^*}(t_1)=\frac{4\sqrt{2}h_{V_1}C_q}{f_{\pi}}\Big\{1-t_1D_{\rho_1}(t_1)\Big\} ,
\end{eqnarray}

\begin{eqnarray}
F_{\omega\eta\gamma^*}(t_1)=\frac{4\sqrt{2}h_{V_1}}{3f_{\pi}}\Big\{C_q\Big(1-t_1D_{\omega_1}(t_1)\Big)
  +\frac{18A^{\eta}_{\phi\omega,1}t_1}{F_{\omega_1}}\Big(D_{\omega_1}(t_1)-D_{\phi_1}(t_1)\Big)\Big\}\, ,
\end{eqnarray}

\begin{eqnarray}
F_{\phi\eta\gamma^*}(t_1)= \frac{4\sqrt{2}h_{V_1}}{3f_{\pi}}
\Bigg\{\Big[2C_s-\Big(\frac{5}{\sqrt{2}}C_q-C_s\Big)A^{\eta}_{1} \Big]
\Big[1-t_1D_{\phi_1}(t_1)\Big]
+\frac{9\sqrt{2}A^{\eta}_{\phi\omega,1}}{F_{\phi_1}}\Big[t_1D_{\omega_1}(t_1)-t_1D_{\phi_1}(t_1)\Big]
\Bigg\} \, ,
\label{phietagam}
\end{eqnarray}
\begin{eqnarray}
 F_{\rho\eta^{\prime}\gamma^*}(t_1)=\frac{4\sqrt{2}h_{V_1}C_q^{\prime}}{f_{\pi}}\Big\{1-t_1D_{\rho_1}(t_1)\Big\} , \ \ F_{\omega\eta^{\prime}\gamma^*}(t_1)=\frac{4\sqrt{2}h_{V_1}C_q^{\prime}}{3f_{\pi}}\Big\{1-t_1D_{\omega_1}(t_1)\Big\}\, ,
\end{eqnarray}
\begin{eqnarray}
F_{\phi\eta^{\prime}\gamma^*}(t_1)=\frac{4\sqrt{2}h_{V_1}}{3f_{\pi}}\Big[-2C_s^{\prime}-
\Big(\frac{5}{\sqrt{2}}C_q^{\prime}+C_s^{\prime}\Big)A_1^{\eta^{\prime}}\Big]
\Big[1-t_1D_{\phi_1}(t_1)\Big]\, .
\label{phietapgam}
\end{eqnarray}

\end{widetext}

\section{Fitting the model parameters to the existing data}\label{fits}
We have fitted the parameters of our model to all existing 
experimental data, which can be described 
  by the Lagrangians Eq.(\ref{l0}-\ref{l7}), in the space-like as well as in the time-like region of
 the photon virtualities.
 The data in the space-like region include measurements of the 
transition form factors for $\pi^0, \eta, \eta^{\prime}$ 
 by BaBar\cite{BABAR:2011ad}, BELLE\cite{Uehara:2012ag}, CELLO \cite{Behrend:1990sr} and CLEO \cite{Gronberg:1997fj} collaborations.
In our model they are predicted in Eqs.(\ref{ffpi0}-\ref{ffetap}). 
The data in the time-like region include measurements of the cross sections for the reactions $e^+ e^- \to \pi^0 (\eta) \gamma $ by SND \cite{Achasov:2006dv,Achasov:2016bfr} and CMD2 \cite{Akhmetshin:2004gw} collaborations. 
 The formula for the $e^+ e^- \to P \gamma $ cross section, where $P$ denotes a pseudoscalar ($\pi^0, \eta$ or $\eta'$), reads
\begin{eqnarray}
\sigma_{e^+e^-\to P\gamma}(s)=
&&\frac{(4\pi\alpha)^3}{24\pi s}
 \Big(1-\frac{m_{P}^2}{s}\Big)\Big(\frac{s-m_P^2}{2\sqrt{s}}\Big)^2 \nonumber \nonumber \\
 && \cdot |F_{\gamma \gamma^*P}(0,s)|^2\, ,
\end{eqnarray}
where $m_P$ is the mass of $P$, $s$ the Mandelstam variable and $F_{\gamma \gamma^*P}(0,s)$ one of the transition form-factors Eqs.(\ref{ffpi0}-\ref{ffetap}).

In addition the time-like form factors measured in 3-body decays 
  were used in the fit.
The following data sets were included and the model parameters
  fitted using the formulae given in brackets: A2 measurement of a decay
   $\pi^0 \to \gamma e^+ e^-$ \cite{Adlarson:2016ykr} (Eq. \ref{ffpi0}),
  A2 measurement of a decay $\eta \to \gamma e^+ e^-$ \cite{Adlarson:2016hpp}
 (Eq. \ref{ffeta}),
BESIII measurement of a decay $\eta^{\prime} \to \gamma e^+ e^-$ \cite{Ablikim:2015wnx}
  (Eq. \ref{ffetap}),
A2 measurement of a decay $\omega \to \pi^0 e^+ e^-$ \cite{Adlarson:2016hpp}
 (Eq. \ref{omegapigam}),
KLOE-2 measurement of a decay $\phi \to \pi^0 e^+ e^-$ \cite{Anastasi:2016qga}
 (Eq. \ref{phipigam}) and
KLOE-2 measurement of a decay $\phi \to \eta e^+ e^-$ \cite{Babusci:2014ldz}
(Eq. \ref{phietagam}). 
For the A2 measurement of a decay $\eta \to \pi^0 \gamma \gamma$ \cite{Nefkens:2014zlt}
 a differential
 partial width was given. The formula describing it reads
\begin{eqnarray}
 d\Gamma\Big(\eta(q) &\to& \pi^0(p) \gamma(k_1) \gamma(k_2)\Big) = \nonumber \\
   &&\frac{1}{4m_{\eta}} |M|^2 dLips_3(q;p,k_1,k_2)\, ,
\end{eqnarray}
with the amplitude given by
\begin{eqnarray}
M &=& \sum_{i,V} \bigg(\frac{4\sqrt{2} e h_{V_i}}{3f_{\pi}} \bigg)^2 \epsilon_{\mu\nu\alpha\beta}
q^{\nu} k_1^{\alpha} \epsilon^{\beta}(k_1) g^{\mu\delta} \nonumber \\ 
&&D_{V_i}((p+k_2)^2)\epsilon_{\delta\sigma{\delta^{\prime}}{\sigma^{\prime}}} p^{\sigma} k_2^{{\delta^{\prime}}} \epsilon^{{\sigma^{\prime}}}(k_2) B_{V_i} \nonumber \\ 
  &&+ \Big( k_1\leftrightarrow k_2 \Big) \, ,
\end{eqnarray}
where $B_{\phi_i}=-\frac{3}{\sqrt{2}} A_i^{\pi^0} [2C_s-(\frac{5}{\sqrt{2}}C_q-C_s)A_i^{\eta}]$, $B_{\rho_i} = 3C_q$, $B_{\omega_i} = 3 H_{\omega_i}C_q$ and $D_{V_i}$ is defined in Eq. (\ref{bwt}).
 
The  2-body partial decay widths \cite{Olive:2016xmw} $P\to\gamma\gamma$
  $V\to e^+e^-$ ($V=\rho,\omega,\phi$), $V\to \pi^0\gamma$, $V\to \eta\gamma$,
 $\phi\to \eta'\gamma$, $\eta'\to \rho\gamma$ and $\eta'\to \omega\gamma$
  were also used in the fits.
 In our model they are expressed as

\begin{equation}
\Gamma(P\to \gamma \gamma)=\frac{m_{P}^3 \pi \alpha^2}{4} 
   |F_{P\gamma^*\gamma^*}(0,0)|^2,
\end{equation}
\begin{equation}
\Gamma(\rho \to e^+ e^-)=\frac{4\pi\alpha^2M_{\rho}f_{V_1}^2}{3},
\end{equation}

\begin{equation}
\Gamma(\omega \to e^+ e^-)=\frac{4\pi\alpha^2M_{\omega}f_{V_1}^2F_{\omega_1}^2}{27},
\end{equation}

\begin{equation}
\Gamma(\phi\to e^+ e^-)=\frac{8\pi\alpha^2M_{\phi}f_{V_1}^2F_{\phi_1}^2}{27},
\end{equation}

\begin{equation}
\Gamma(P\to V \gamma)=\frac{\alpha}{8} m_{P}^3 k_V^3 |F_{VP\gamma^*}(0)|^2,
\end{equation}

\begin{equation}
\Gamma(V \to P \gamma)=\frac{\alpha}{24} M_{V}^3k_{\mathcal{P}}^3|F_{VP\gamma^*}(0)|^2,
\end{equation}
where $k_V=(1-\frac{m_P^2}{M_{V}^2}\Big)$, 
 $k_P=(1-\frac{M_{V}^2}{m_{P}^2}\Big)$.
The form factors $F_{P\gamma^*\gamma^*}$  are given
   in Eqs.(\ref{ffpi0}-\ref{ffetap})
  and the form factors $F_{VP\gamma^*}$ are given in Eqs.(\ref{rhopigam}-\ref{phietapgam}).

\begin{figure}
\begin{center}
\hskip-1cm\includegraphics[width=9.cm]{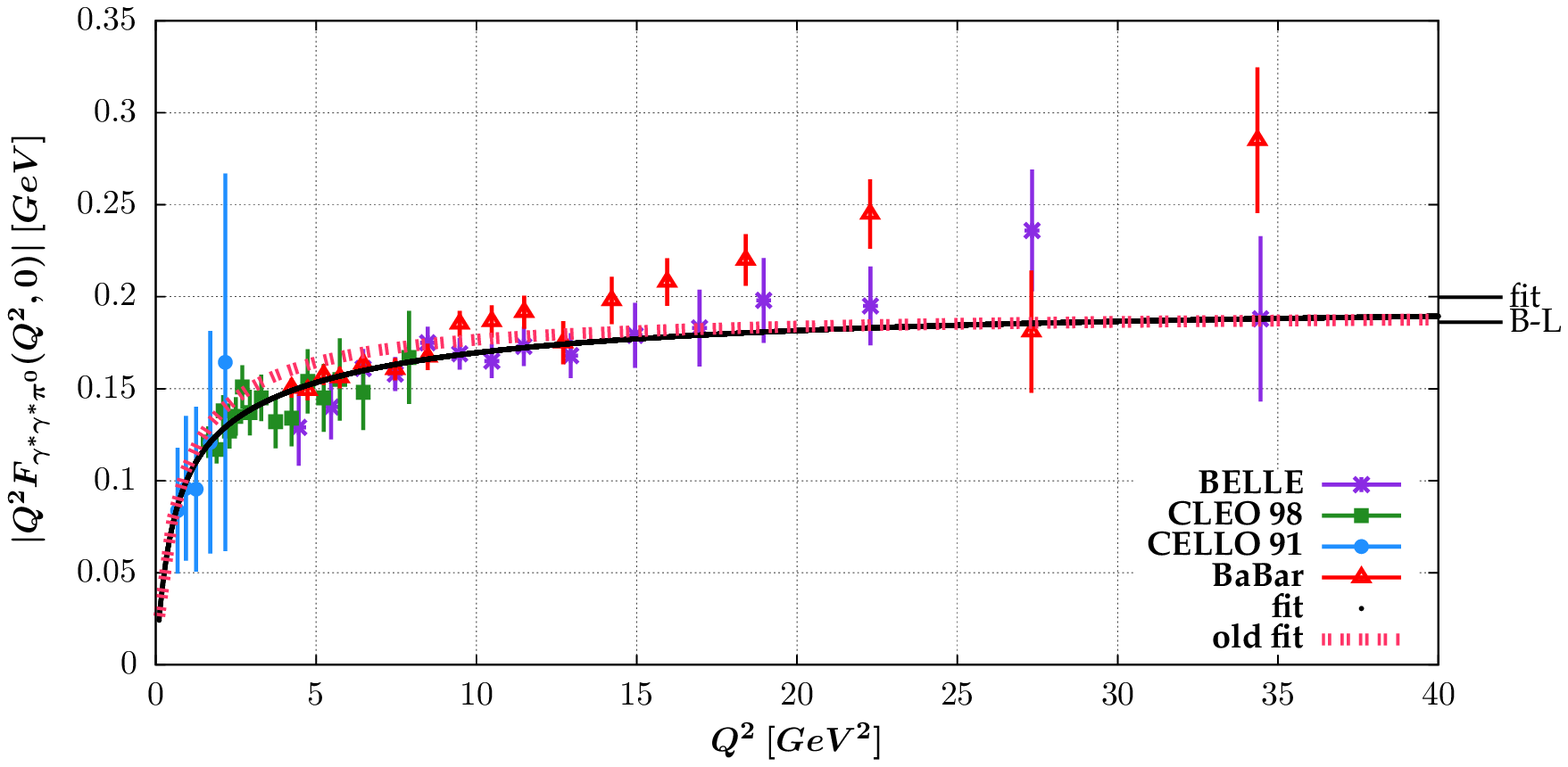}
\phantom{}\hskip-1cm\includegraphics[width=9.cm]{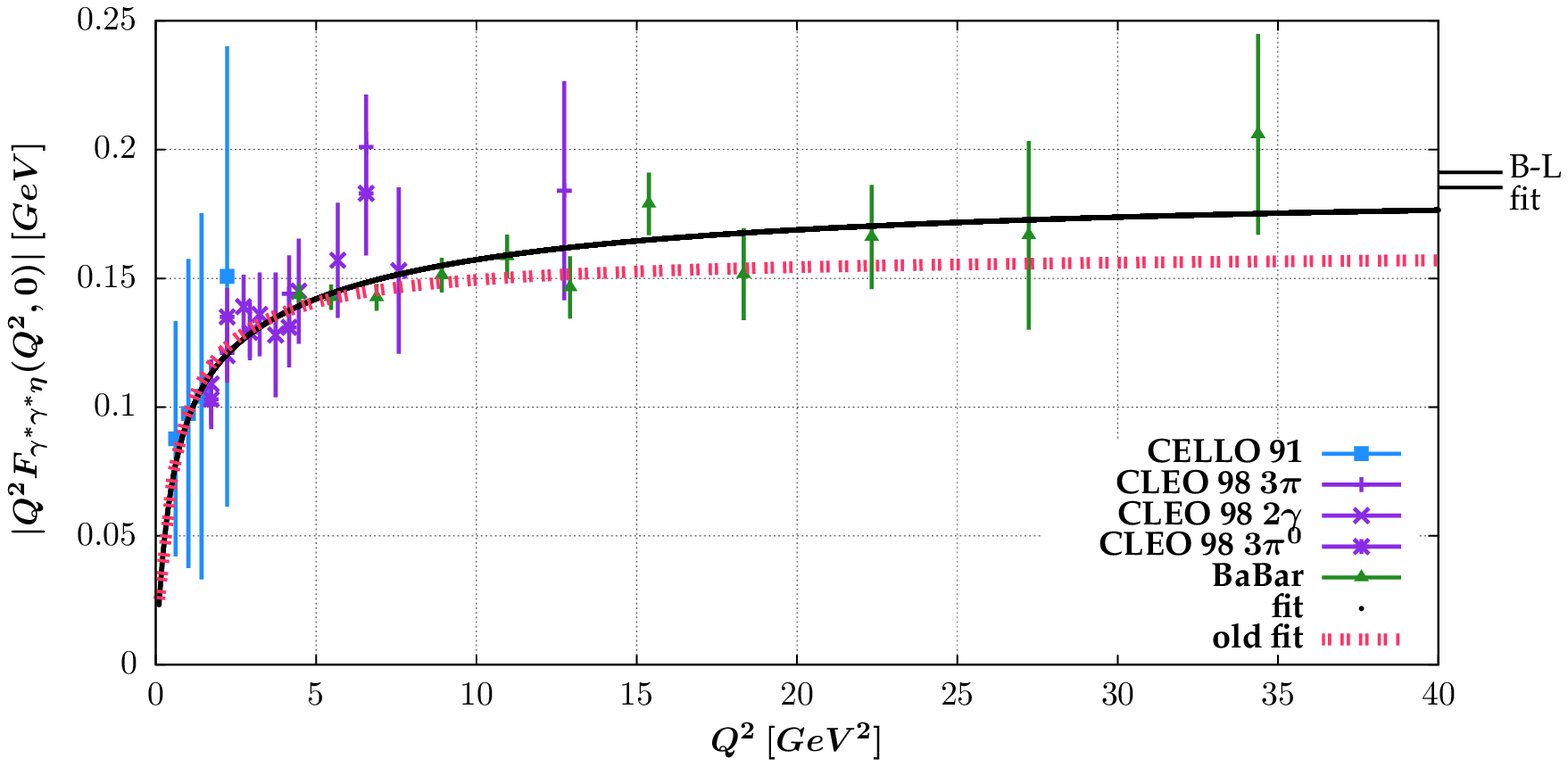}
\phantom{}\hskip-1cm\includegraphics[width=9.cm]{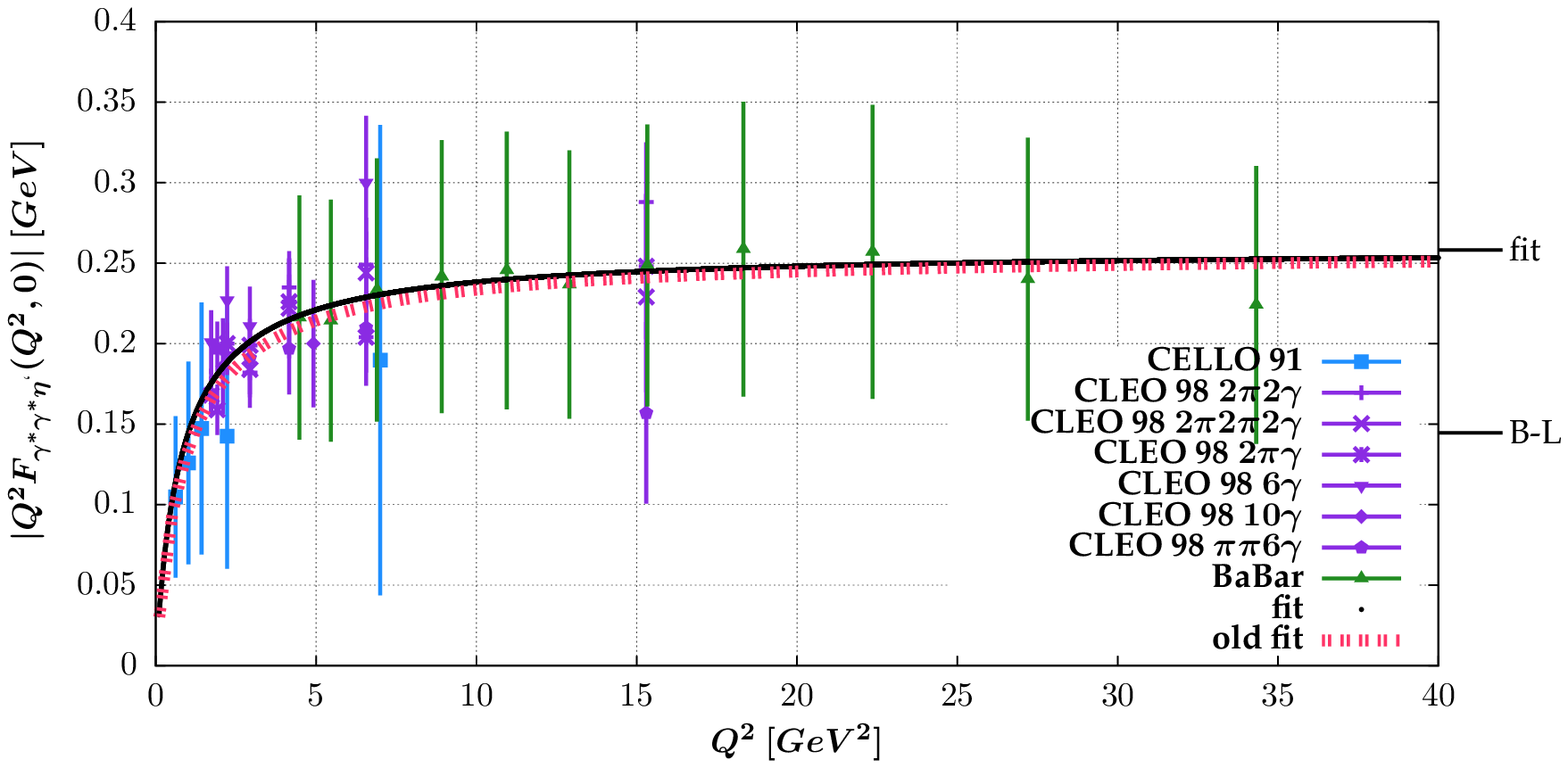}
\caption{Transition form factors $\gamma^* \gamma P$ in the space-like region
 compared to the data.
\label{plot1}}
\end{center}
\end{figure}

\begin{figure}
\begin{center}
\hskip-1.cm\includegraphics[width=9.6cm]{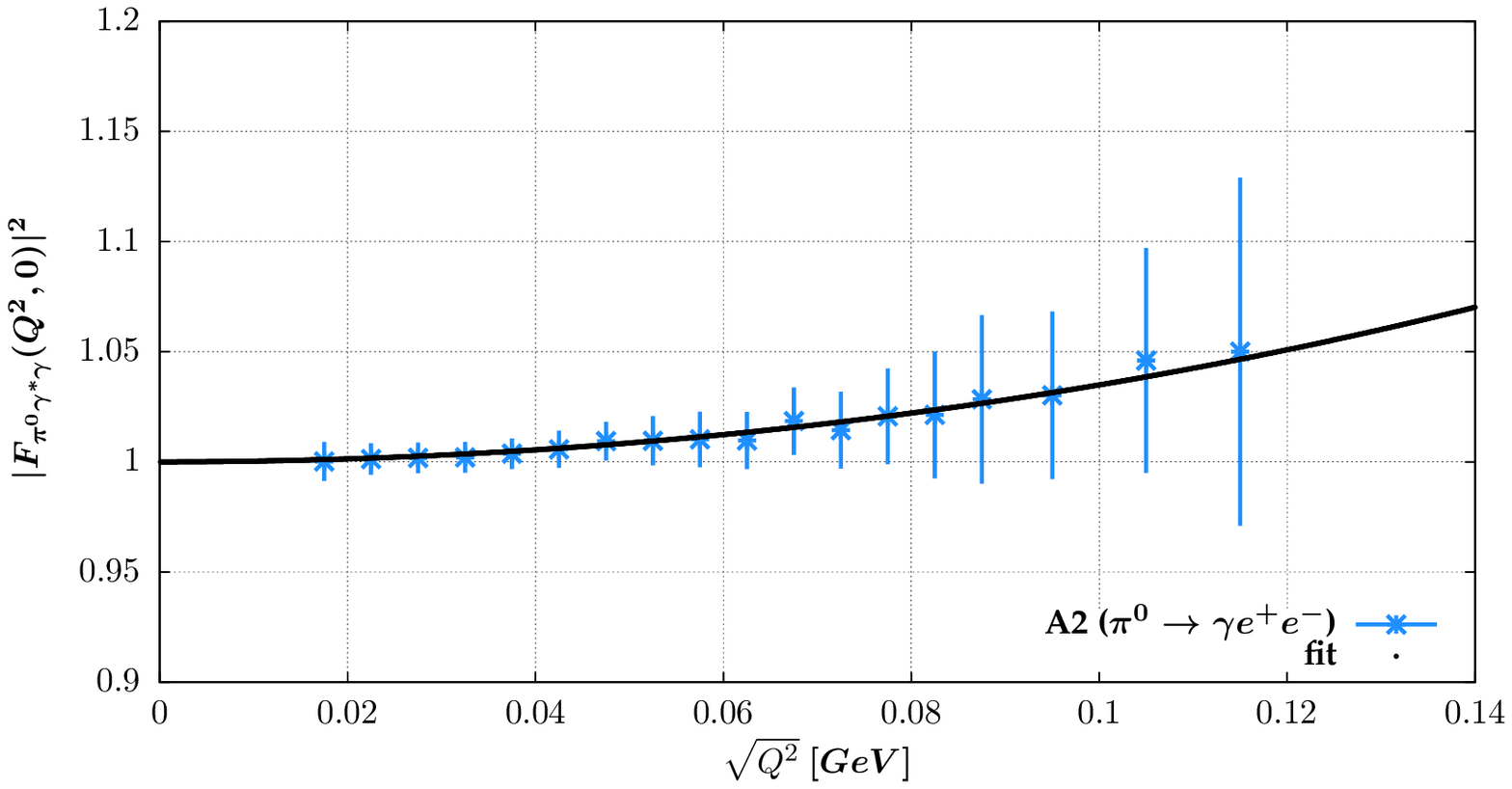}
\phantom{}\hskip-0.6cm\includegraphics[width=9.2cm]{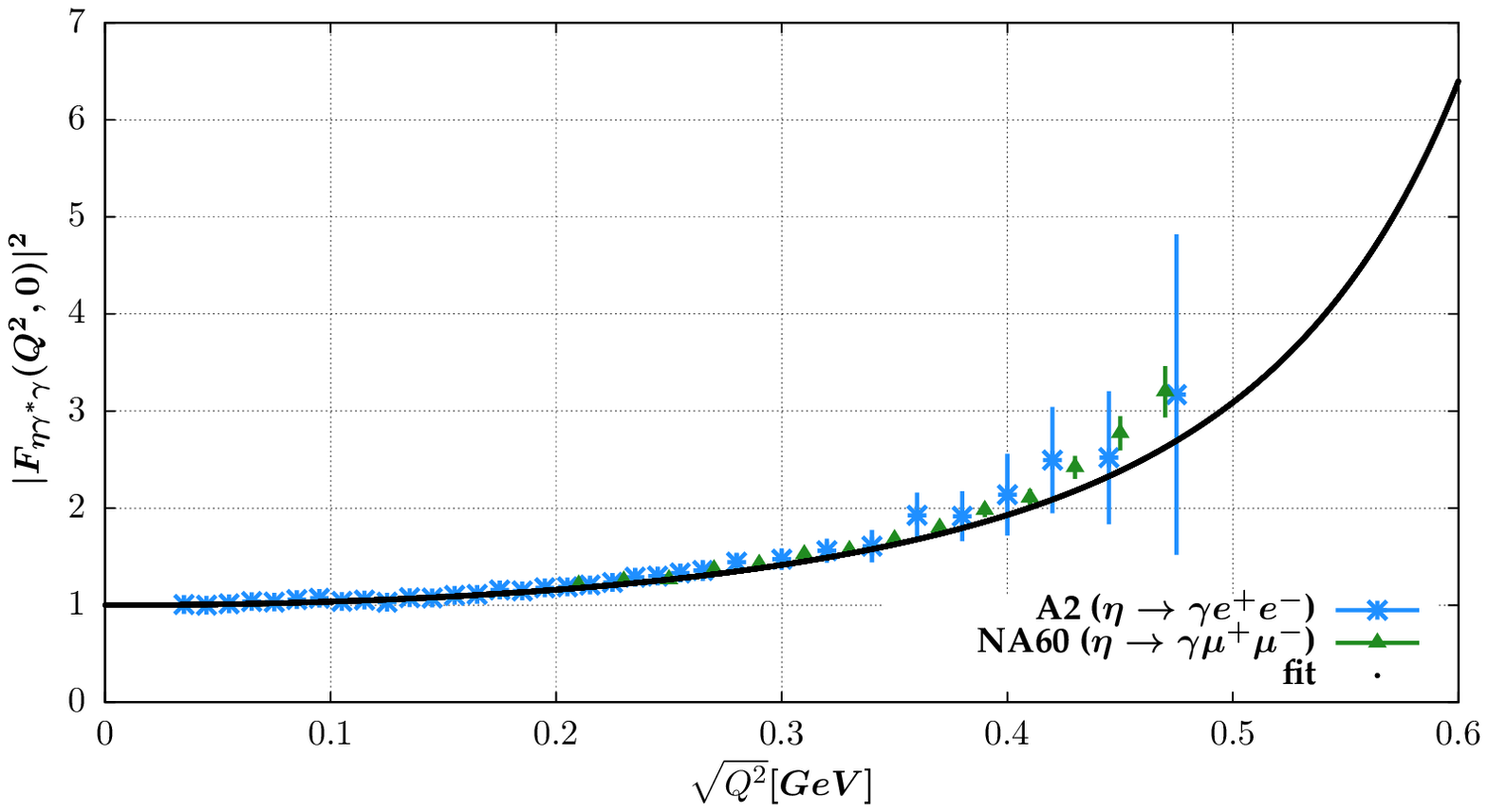}\hskip 1.6cm
\phantom{} \hskip-1.7cm\includegraphics[width=8.8cm]{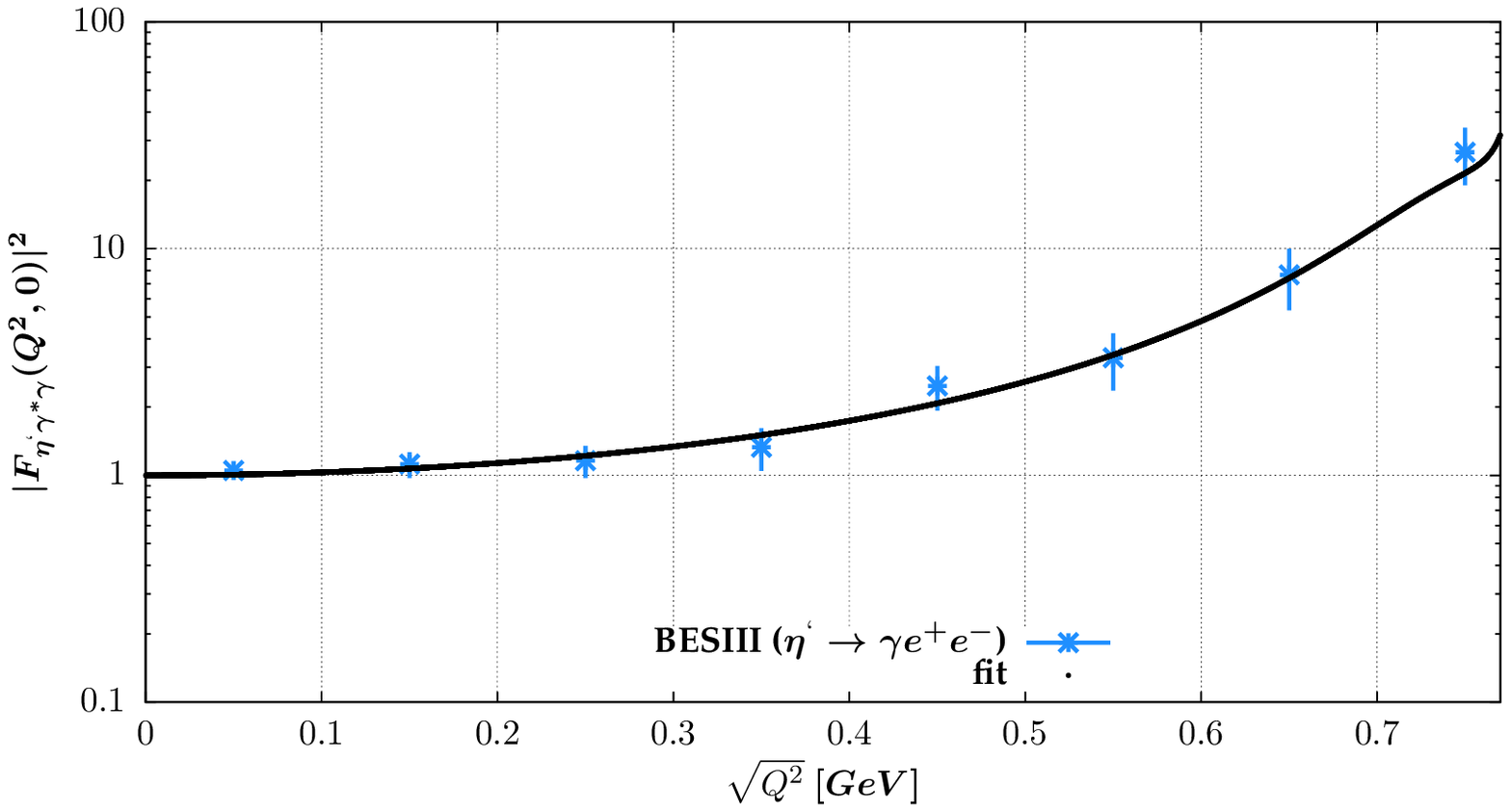}
\caption{Transition form factors $\gamma^* \gamma P$ in the time-like region
 compared to the data.
\label{plot10}
}
\end{center}
\end{figure}

\begin{figure}
\begin{center}
\hskip-1cm\includegraphics[width=9.cm]{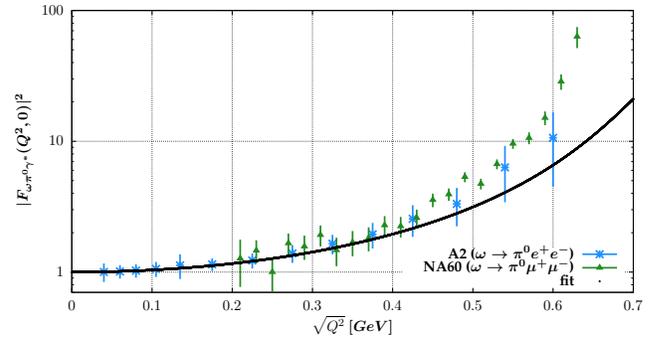}
\caption{The form factor $\omega\pi^0 \gamma $ in the time-like region
 compared to the data.
\label{plot13}
}
\end{center}
\end{figure}

We have performed two fits. One with fixed parameters $\theta_8$, $\theta_0$, $f_8$, $f_0$ and $f_\pi$ describing the  $\eta-\eta^{\prime}$ mixing and the 
 $\pi^0\to\gamma\gamma$ decay width (called fit 1) and the second one where we fit also these parameters (called fit 2) . The $\chi^2$ values for all the 
 experimental sets of data obtained
 in the fits  
 are given in Table \ref{tab1}. BaBar measurement of the  $\pi^0$ transition form factor  
  \cite{Aubert:2009mc} as well as NA60 measurements \cite{Arnaldi:2016pzu} 
 of the $\eta$ transition form factor
  and the $F_{\omega\pi^0\gamma^*}$ form factor were not used in the fits summarized here.
  They are in contradiction with other experimental data (see Figures \ref{plot1},\ref{plot10} and \ref{plot13}).
  The smallest tension is between
  the  $\eta$ transition form factor measurements of A2 \cite{Adlarson:2016hpp} and
 NA60 \cite{Arnaldi:2016pzu} (see Figure \ref{plot10}) and in fact the data are consistent
  within the experimental error bars. 
  Yet, within the model we developed here, there is no way
  to fit simultaneously SND \cite{Achasov:2006dv} data on $e^+e^- \to \eta\gamma$ cross section,
 the differential width $(\eta \to \pi^0 \gamma \gamma)$ measured by A2 \cite{Nefkens:2014zlt}
 and the partial widths $V\to\eta\gamma$ \cite{Olive:2016xmw} together with
   the  NA60 measurements \cite{Arnaldi:2016pzu} of the $\eta$ transition form factor in
   the time-like region.
\begin{widetext}
\begin{table} [ht]
\begin{center}
\vskip0.3cm
\begin{tabular}{|c|c|c|c|c|c|c|c|}
\hline
 Experiment & nep & $\chi^2$,fit 1 & $\chi^2$,fit 2 & Experiment & nep & $\chi^2$,fit 1 & $\chi^2$,fit 2\\ 
\hline
space-like form-factors & &  & &  & &  &\\
\hline 
BELLE ($\pi^0$)\cite{Uehara:2012ag}          & 15 & 9.96  & 6.72& CLEO98($\eta$) \cite{Gronberg:1997fj}        & 19 & 15.8  & 15.5 \\
CELLO91($\pi^0$) \cite{Behrend:1990sr}       & 5 & 0.34 & 0.24&BaBar($\eta^{\prime} $) \cite{BABAR:2011ad}    & 11 & 5.4 & 3.70\\
CLEO98($\pi^0$) \cite{Gronberg:1997fj}       & 15 & 10.6  & 6.82&CELLO91($\eta^{\prime}$) \cite{Behrend:1990sr} & 5 & 0.73  & 0.56 \\
BaBar($\eta$) \cite{BABAR:2011ad}            & 11 & 7.34 & 7.5&CLEO98($\eta^{\prime} $) \cite{Gronberg:1997fj}& 29 & 25.1   & 24.4\\
CELLO91($\eta $) \cite{Behrend:1990sr}       & 4 & 0.16 & 0.16 & & & & \\
\hline
 $e^+e^-$ cross sections & & & & & & & \\
\hline 
CMD2($\pi^0\gamma$) \cite{Akhmetshin:2004gw} & 46 & 54.1 & 54.1&SND($\eta \gamma $) \cite{Achasov:2006dv}    & 78 & 68.7  & 59.8 \\
SND($\pi^0\gamma$) \cite{Achasov:2016bfr}    & 62 & 65.5 & 54.2&BaBar($\eta \gamma,\eta' \gamma$) \cite{Aubert:2006cy} & 2 & 0.18 & 1.57\\
CMD2 ($\eta \gamma$) \cite{Akhmetshin:2004gw}& 42 & 25.4 & 25.6 & & & & \\
\hline
 3-body decays & &  & &  & &  &\\
\hline 
A2($\pi^0 \to \gamma e^+ e^-$) \cite{Adlarson:2016ykr}        & 18& 0.32 & 0.34&A2($\omega \to \pi^0 e^+ e^-$)	\cite{Adlarson:2016hpp}	      & 14& 2.14  & 2.12 \\
A2($\eta \to \gamma e^+ e^-$) \cite{Adlarson:2016hpp}         & 34& 10.2  & 11.1&KLOE-2($\phi \to \pi^0 e^+ e^-$) \cite{Anastasi:2016qga}       & 15& 4.33 & 4.33 \\
A2 $(\eta \to \pi^0 \gamma \gamma)$ \cite{Nefkens:2014zlt}    & 7& 26.6  & 19.5&KLOE-2($\phi \to \eta e^+ e^-$) \cite{Babusci:2014ldz}	      & 92& 95.1 & 95.1\\
BESIII($\eta^{\prime} \to \gamma e^+ e^-$)\cite{Ablikim:2015wnx}& 8& 2.39 & 2.13 &&&&\\
\hline
 2-body decays & &  & &  & &  &\\
\hline 
$\Gamma(\pi^0 \to \gamma \gamma)$ \cite{Olive:2016xmw}& 1 & 0.36  & 0.1&$\Gamma(\rho \to \pi^0 \gamma)$ \cite{Olive:2016xmw}& 1 & 1.17  & 0.42\\
$\Gamma(\eta \to \gamma \gamma)$ \cite{Olive:2016xmw}& 1 & 0.78 & 2.73&$\Gamma(\omega \to \pi^0 \gamma)$ \cite{Olive:2016xmw}& 1& 4.08  & 1.56\\
$\Gamma(\eta^{\prime} \to \gamma \gamma)$ \cite{Olive:2016xmw}& 1 & 1.05 & 0.44&$\Gamma(\phi \to \pi^0 \gamma)$ \cite{Olive:2016xmw}& 1 & 0.08 & 0.06\\
$\Gamma(\eta^{\prime} \to \rho \gamma)$ \cite{Olive:2016xmw}& 1 & 3.0 & 0.77&$\Gamma(\rho \to \eta \gamma)$ \cite{Olive:2016xmw}& 1 	& 3.32 & 6.8\\
$\Gamma(\eta^{\prime} \to \omega \gamma)$ \cite{Olive:2016xmw}& 1 & 0.00  & 0.54&$\Gamma(\omega \to \eta \gamma)$ \cite{Olive:2016xmw}& 1 & 6.86  & 3.04\\
$\Gamma(\rho \to e^+ e^-)$ \cite{Olive:2016xmw}& 1 & 0.23 &  0.05&$\Gamma(\phi \to \eta \gamma)$ \cite{Olive:2016xmw} & 1 & 1.63  & 1.17\\
$\Gamma(\omega \to e^+ e^-)$ \cite{Olive:2016xmw}& 1 & 0.56 & 0.73& $\Gamma(\phi \to \eta^{\prime} \gamma)$ \cite{Olive:2016xmw}& 1 & 0.01  & 0.00 \\
$\Gamma(\phi \to e^+ e^-)$ \cite{Olive:2016xmw}& 1 & 0.69  &  0.46 & & & &\\
\hline
 & & & & Total& 536 & 454 & 415 \\
\hline
\end{tabular}
\caption{The values of the $\chi^2$ for the experiments used in the fits described in the text. 'nep' means number of experimental points.}
\label{tab1}
\end{center}
\end{table}
\end{widetext}

 In Table \ref{tab2} we give the parameters obtained in both fits. The fit is much better if we allow
 for changing of the $\eta-\eta'$ mixing parameters. In principle one can think of the 'fit 2' as
 a way to extract the $\eta-\eta'$ mixing parameters. Yet, one has to remember 
   that this is a model dependent extraction.

\begin{figure}
\begin{center}
\hskip-1cm\includegraphics[width=9.cm]{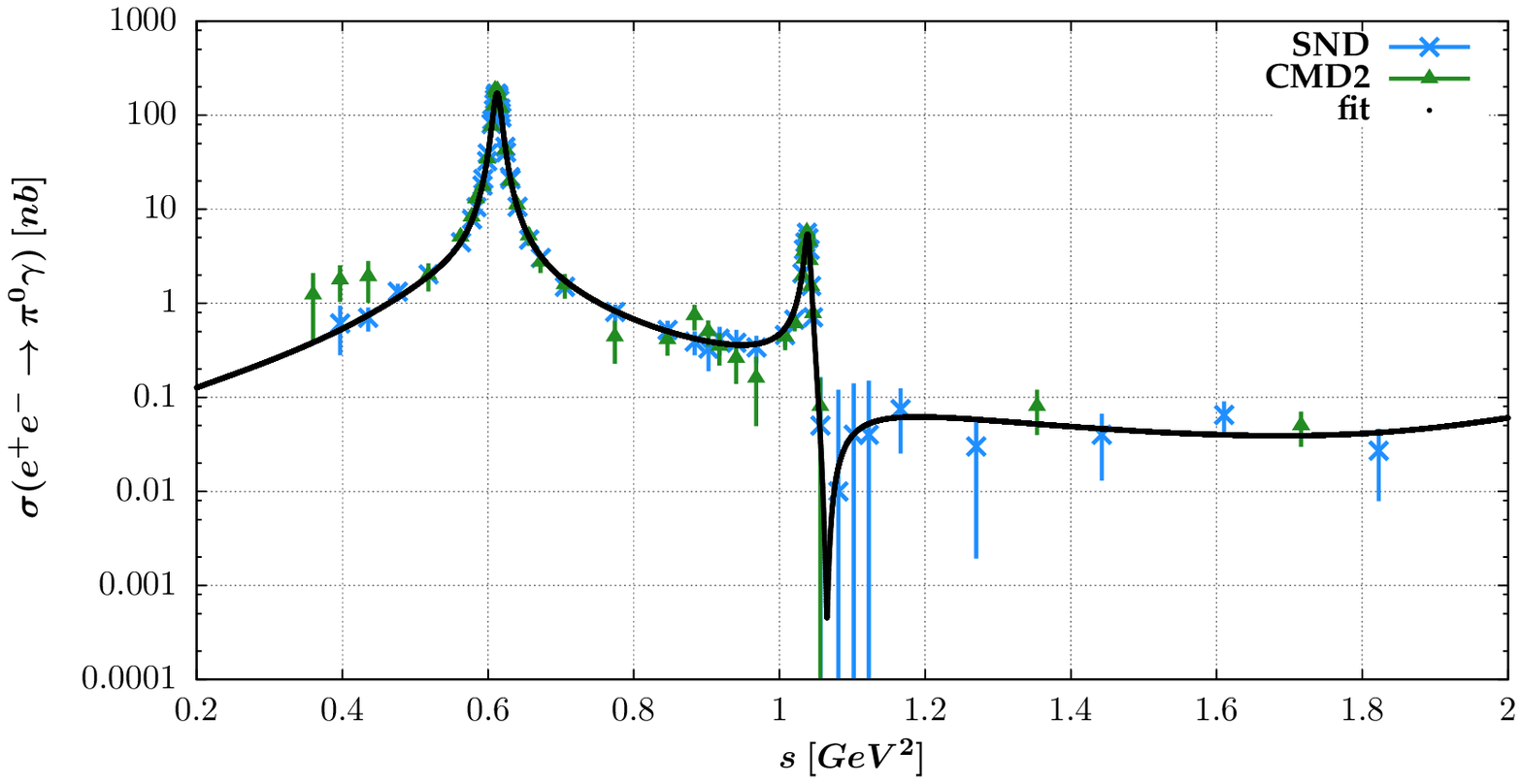}
\phantom{}\hskip-1cm\includegraphics[width=9.cm]{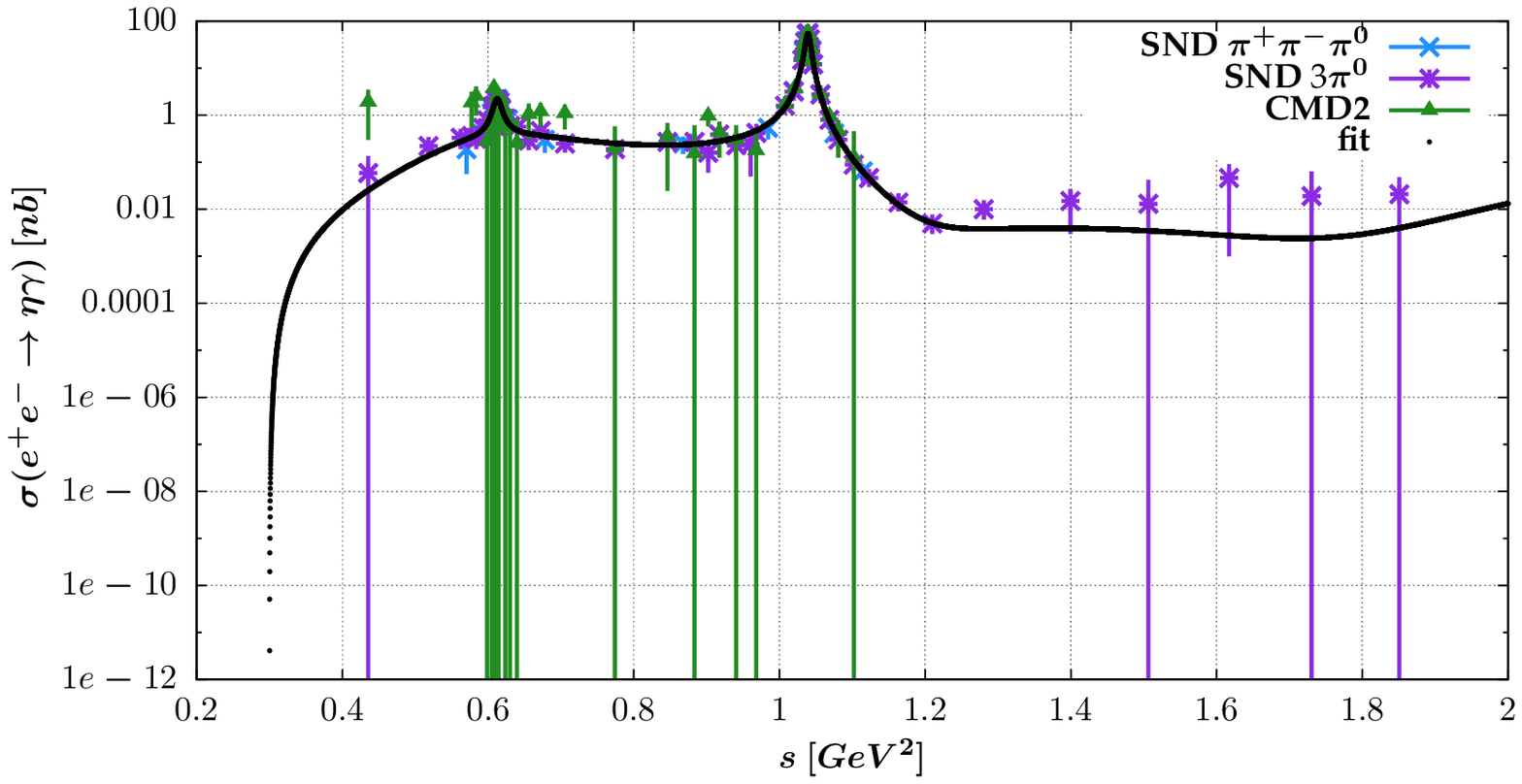}
\caption{Experimental data for $\sigma(e^+ e^- \to P \gamma)$ compared to the model predictions. 
\label{plot4}
}
\end{center}
\end{figure}

\begin{figure}
\begin{center}
\hskip-1cm\includegraphics[width=9.cm]{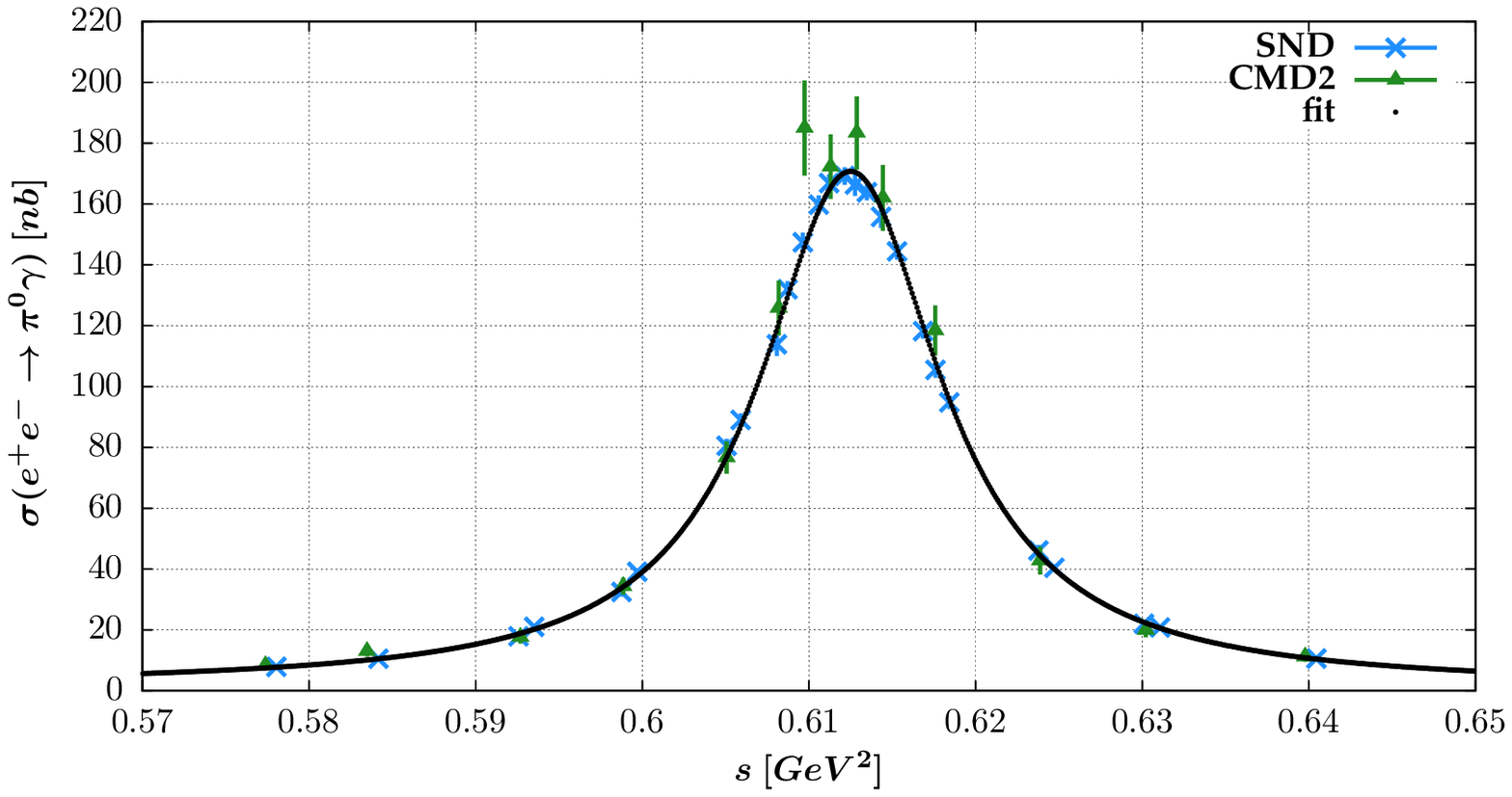}
\phantom{}\hskip-1cm\includegraphics[width=9.cm]{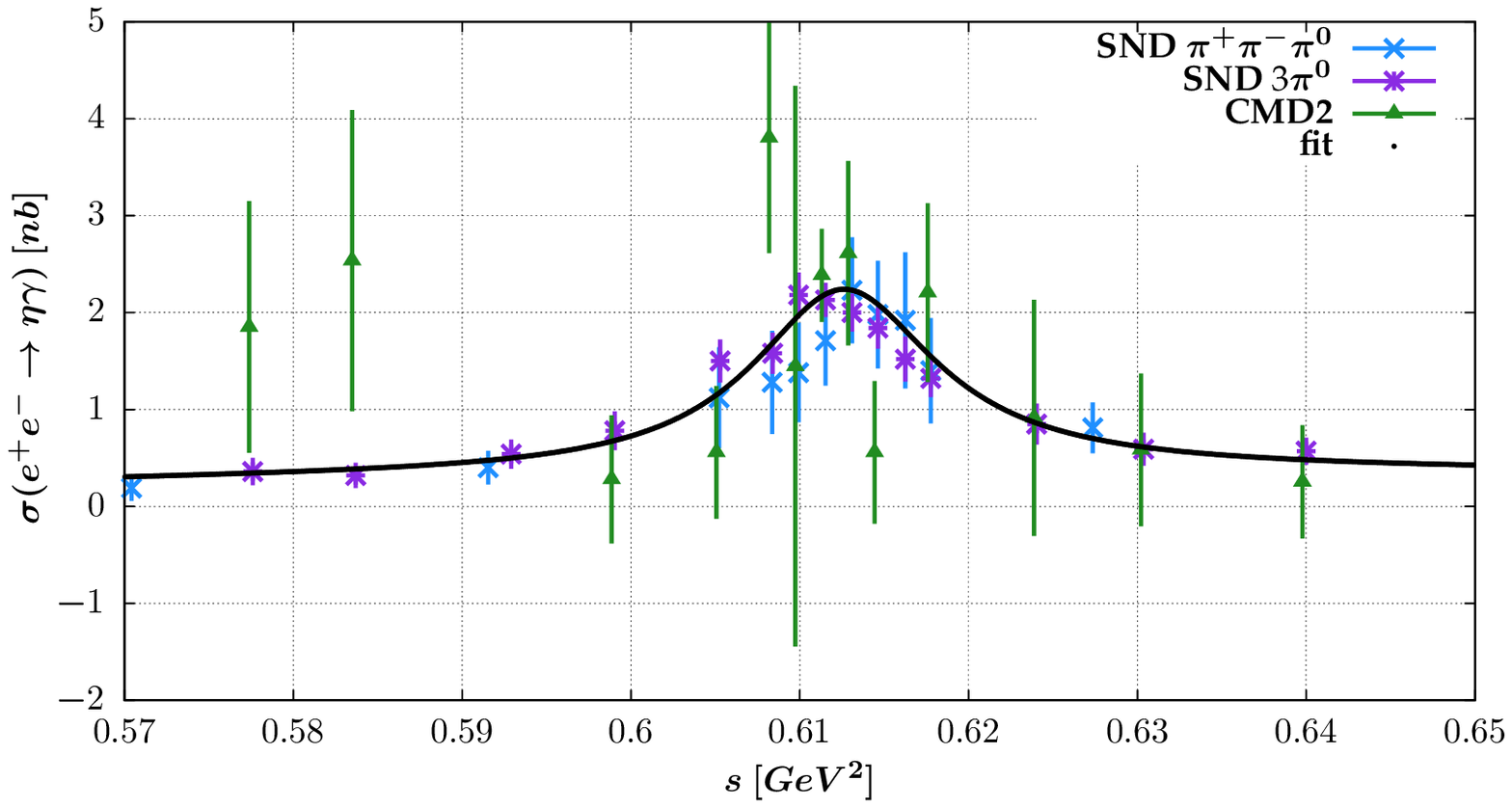}
\caption{Experimental data for $\sigma(e^+ e^- \to P \gamma)$ compared to the model predictions. The region of the $s$ has been limited to $\omega$ peak.
\label{plot5}
}
\end{center}
\end{figure}

\begin{figure}
\begin{center}
\hskip-1cm\includegraphics[width=9.cm]{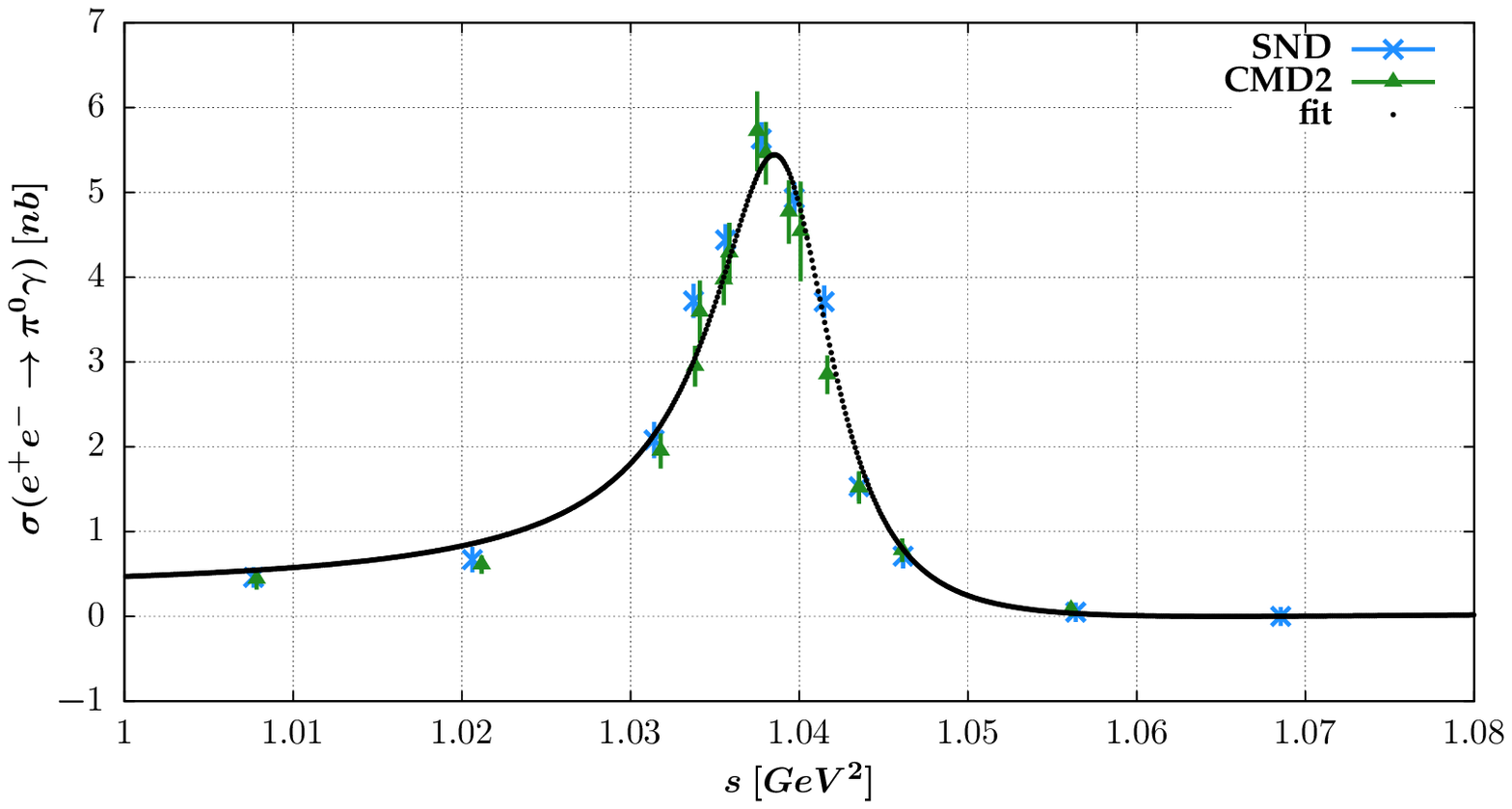}
\phantom{}\hskip-1cm\includegraphics[width=9.cm]{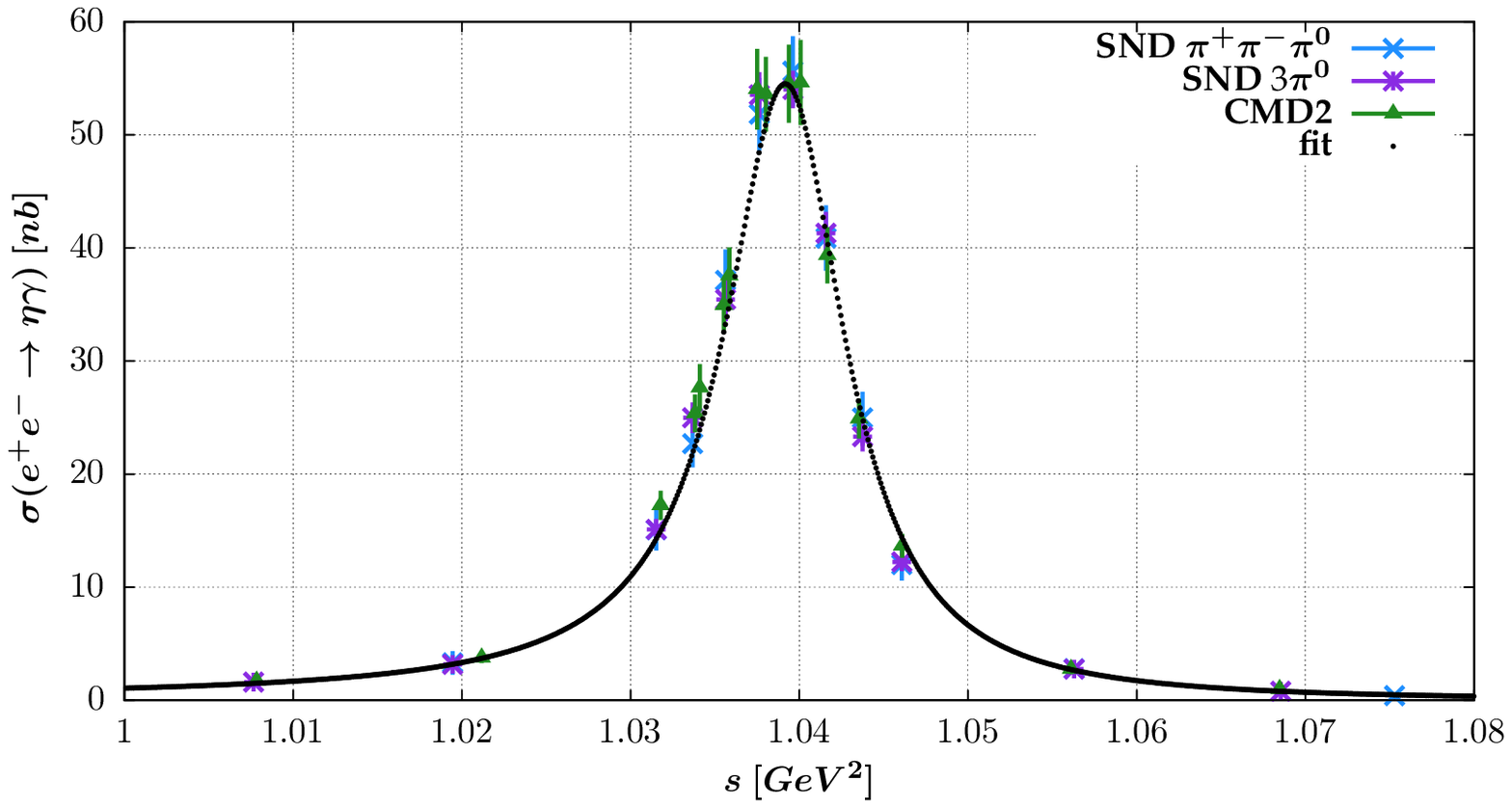}
\caption{Experimental data for $\sigma(e^+ e^- \to P \gamma)$  compared to the model predictions. The region of the $s$ has been limited to $\phi$ peak.
\label{plot6}
}
\end{center}
\end{figure}

\begin{figure}
\begin{center}
\hskip-1cm\includegraphics[width=9.cm]{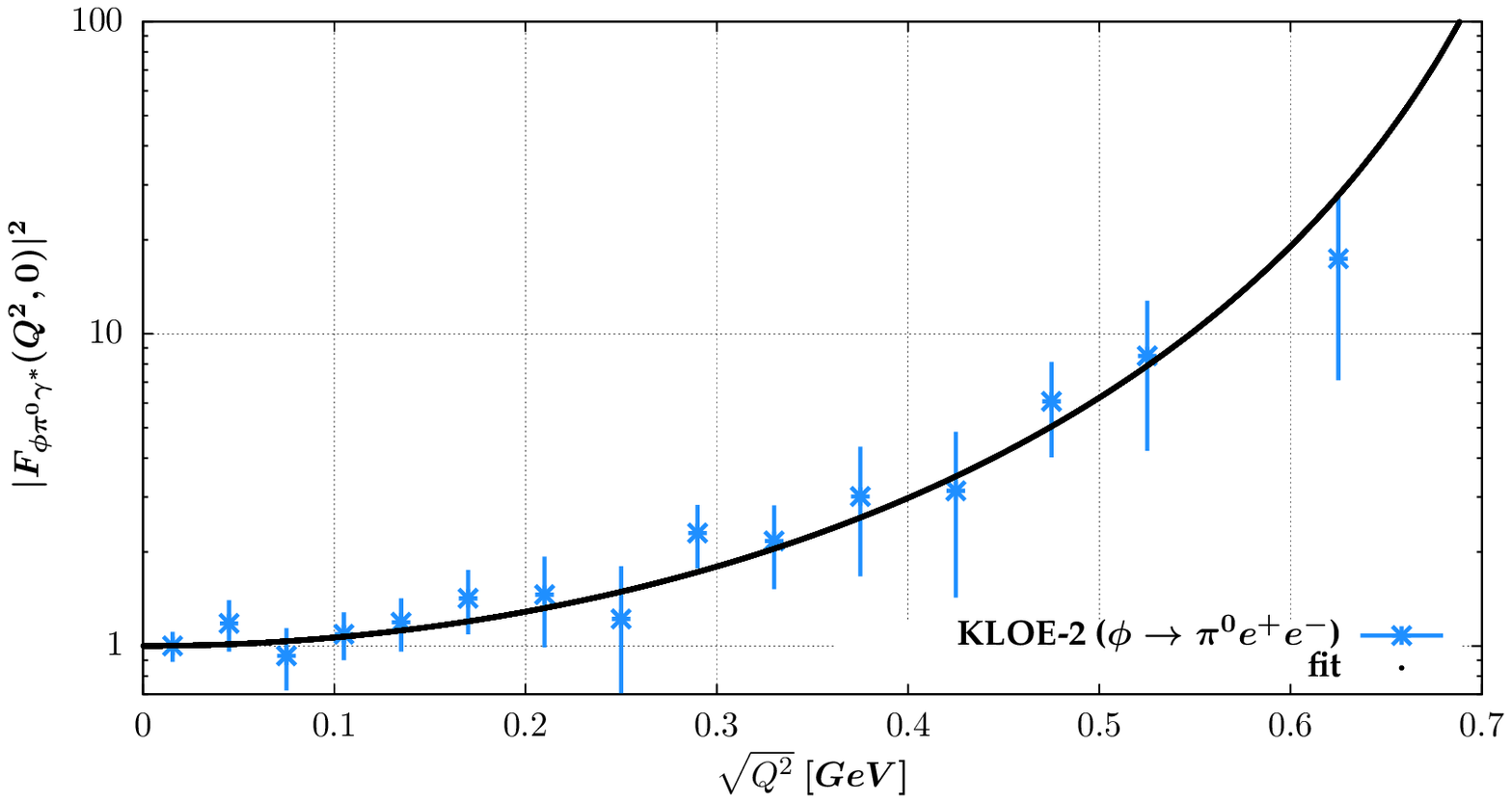}
\phantom{}\hskip-1cm\includegraphics[width=9.cm]{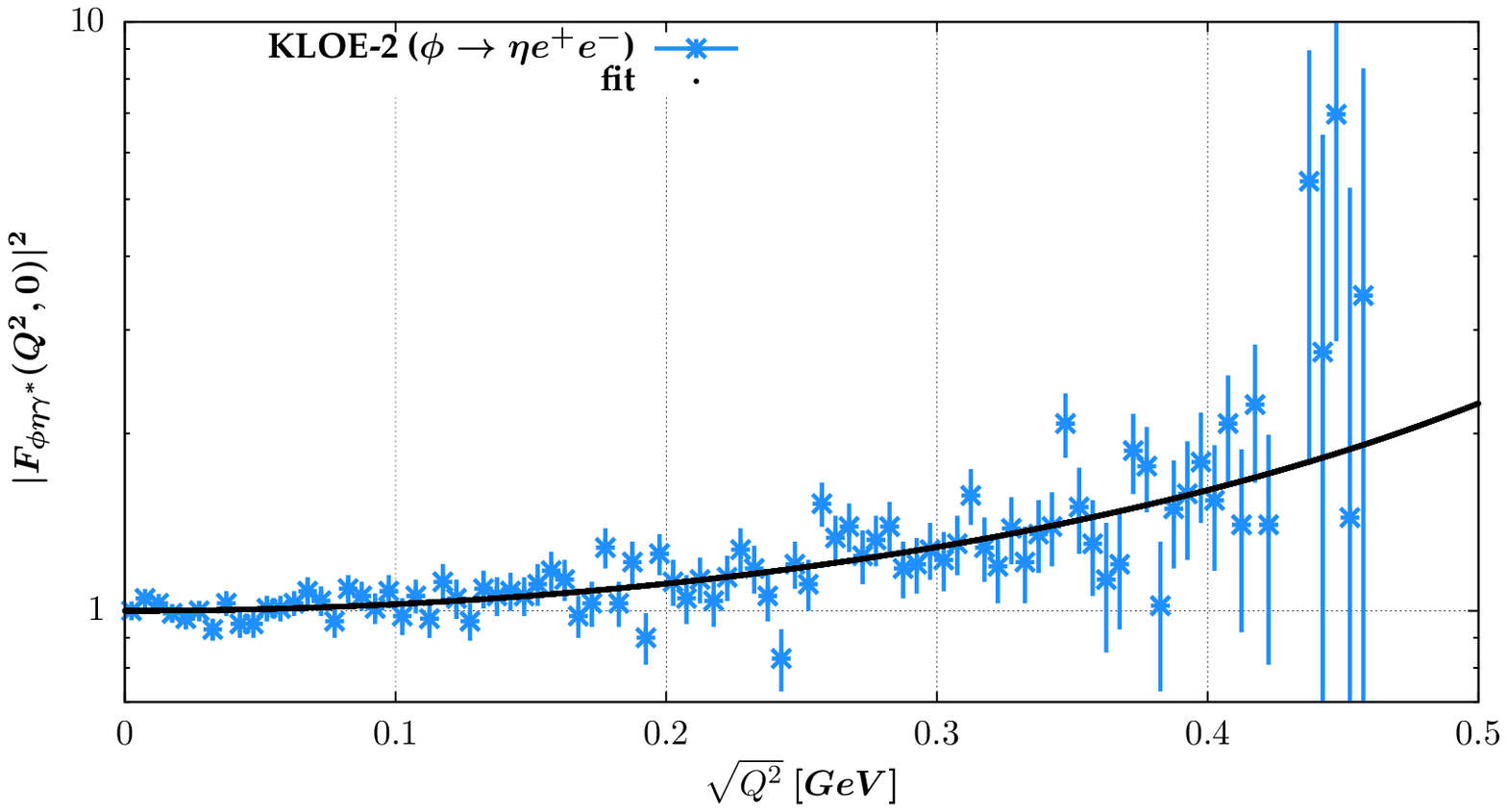}
\caption{The form factor $\phi P \gamma $ in the time-like region
 compared to the data.
\label{plot14}
}
\end{center}
\end{figure}

\begin{figure}
\begin{center}
\hskip-1cm\includegraphics[width=9.cm]{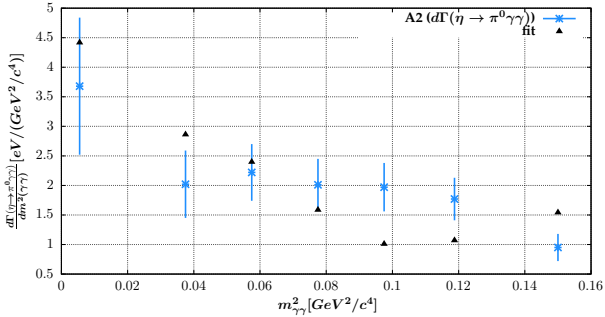}
\caption{The differential partial width of the decay $\eta\to\pi^0\gamma\gamma$
 compared to the data.
\label{plot16}
}
\end{center}
\end{figure}

To show how the fits represent data for individual data points we present here the following plots:
\begin{itemize}
 \item{In  Figure \ref{plot1} the pseudoscalars transition form factors 
 in the space-like region are presented. The 'old fit' refers there to the 2-octet model from \cite{Czyz:2012nq}. On the right-hand side of the plots the asymptotic values of the form factors are given within the current model
 (fit 2) (see also discussion in Section \ref{secas}) and as in original Brodsky-Lapage paper \cite{Lepage:1980fj} i.e. $2f_\pi$ for the pion form factor, 
 $2f_\eta = 2f_\pi/(\frac{5}{3}C_q-\frac{\sqrt{2}}{3}C_s)$ for the eta form factor
 and $2f_{\eta'} = 2f_\pi/(\frac{5}{3}C^{\prime}_q+\frac{\sqrt{2}}{3}C^{\prime}_s)$ for the eta prime form factor}
\item{In  Figures \ref{plot4}-\ref{plot6} the cross sections of the reactions
  $e^+ e^- \to \pi^0 \gamma$ and $e^+ e^- \to \eta \gamma$ are shown. We show all the data points
  and fits
 in Figure \ref{plot4}  and separately show the regions around $\omega$
 ( Fig. \ref{plot5} ) and $\phi$ ( Fig. \ref{plot6}) resonances.}
\item{In Figure \ref{plot10} the pseudoscalars transition form factors 
 in the time-like region are presented.}
\item{In Figures \ref{plot13}-\ref{plot14} the $VP\gamma$ form factors are shown.}
\item{In Figure \ref{plot16} the differential decay width of $\eta\to\pi^0\gamma\gamma$ decay is
  presented.}
\end{itemize}

We show only the plots using the parameters from fit~2. The plots with
  the fit 1 parameters look similar.

\begin{table}[hb]
\begin{center}
\vskip0.3cm
\begin{tabular}{|c|c|c|}
\hline
 Parameter & fit 1 & fit 2  \\ 
\hline
$h_{V_1}$               & 0.0335(2)  & 0.0377(8) \\
$f_{V_1}$                & 0.2022(8)   & 0.2020(8) \\
$f_{V_2}h_{V_2}$         & -0.0013(2)  & -0.0010(4)\\
$h_{V_2}$               &  0.00184(5) & 0.0002(1)\\
$h_{V_3}$               &  -0.485(7) & -0.30(4)\\
$H_{\omega_1}$          &  1.160(11)   &  1.02(3)\\
$F_{\omega_1}$           & 0.881(8)    & 0.88(1)\\      
$F_{\phi_1}$     	& 0.783(5)     & 0.783(5) \\
$A_1^{\pi^0}$           &  -0.094(1)   & -0.083(2)\\
$A_2^{\pi^0}$            & -12.04(16)  & -15(6)\\
$A_3^{\pi^0}$            &  0.08(3)  & -0.16(7)\\
$A_1^{\eta}$            &  -0.041(4)   & -0.30(4)\\
$A_3^{\eta}$            &  0.23(6)   & -0.06(8)\\
$A_1^{\eta^{'}}$        &   -0.039(7)  & -0.21(5)\\ 
$A_3^{\eta^{'}}$        &   -0.27(3)  & -0.56(6)\\ 
$A^{\pi^0}_{\phi\omega,1}$&  -0.23(4)    & -0.21(4)\\
$A^{\eta}_{\phi\omega,1}$  & -0.031(8)  & -0.028(7)\\
$f_\pi$               & 0.092388(f)  & 0.09266(8)\\
$f_0$               & 0.10623(f)  & 0.095(2)\\
$f_8$               & 0.11697(f)  & 0.17(1)\\
$\theta_0$               & -0.14471(f)  & -0.54(12)\\
$\theta_8$               & -0.36516(f)  & -0.446(17)\\
\hline
\end{tabular}
\caption{Model parameters obtained in the fits. The errors,
 given in brackets, are the parabolic errors calculated by Minos of the 
 Minuit package.(f) means that the parameter was fixed in the fit to the 
  value given in this Table.}
\label{tab2}
\end{center}
\end{table}


\section{The asymptotics of the form factors and slopes of the form factors at the origin}\label{secas}
\label{asymp}

 The analytic form of the asymptotic behaviour of the form factors
 is analogous to the one obtained in \cite{Czyz:2012nq} with the 
  asymptotic limits changed. For completeness we report here the
  formulae, but skip the discussion as it should repeat the one presented 
  in \cite{Czyz:2012nq}. They read  

\begin{eqnarray}
 &&\kern-20pt F_{\gamma^*\gamma\pi^0}(t,0) = \sum^n_{i=1}\frac{4\sqrt{2}h_{V_i}f_{V_i}}{3f_{\pi}} \frac{1}{t}  \nonumber \\
 &&\kern-10pt\bigg(M^2_{\rho_i} + F_{\omega_i}H_{\omega_i}M^2_{\omega_i}+A_i^{\pi^0}F_{\phi_i}M^2_{\phi_i}\bigg) + O\bigg(\frac{1}{t^2}\bigg)\, ,
\end{eqnarray}

\begin{eqnarray}
 &&\kern-10pt F_{\gamma^*\gamma^*\pi^0}(t,t) = \sum^n_{i=1}\frac{4\sqrt{2}h_{V_i}f_{V_i}}{3f_{\pi}} \frac{1}{t^2} \nonumber \\ 
 &&\bigg(-2M^2_{\rho_i}M^2_{\omega_i} - (A_i^{\pi^0}F_{\phi_i}-A_{\phi\omega,i}^{\pi^0})M^4_{\phi_i} 
 - 2A_{\phi\omega,i}^{\pi^0}M^2_{\phi_i}M^2_{\omega_i} \nonumber \\ 
 &&- (F_{\omega_i}H_{\omega_i}-1-A_{\phi\omega,i}^{\pi^0}) M^4_{\omega_i}\bigg) + O\bigg(\frac{1}{t^3}\bigg)\, ,
\end{eqnarray}

\begin{eqnarray}
 &&\kern-30pt F_{\gamma^*\gamma\eta}(t,0) = \sum^n_{i=1}\frac{4\sqrt{2}h_{V_i}f_{V_i}}{3f_{\pi}} \frac{1}{t} \nonumber\\ 
 &&\bigg(3C_qM^2_{\rho_i}+\frac{1}{3}F_{\omega_i}C_qM^2_{\omega_i} 
  - \frac{2\sqrt{2}}{3}C_sF_{\phi_i}M^2_{\phi_i} \nonumber \\ 
 && + \big(\frac{5}{3}C_q-\frac{\sqrt{2}}{3}C_s\big)A_i^{\eta}F_{\phi_i}M^2_{\phi_i} \bigg) + O\bigg(\frac{1}{t^2}\bigg)\, ,
\end{eqnarray}

\begin{eqnarray}
 &&\kern-25ptF_{\gamma^*\gamma^*\eta}(t,t) = \sum^n_{i=1}\frac{8\sqrt{2}h_{V_i}f_{V_i}}{f_{\pi}}\frac{1}{t^2} \nonumber \\ 
 &&\kern-10pt\bigg( -\frac{1}{2}C_qM^4_{\rho_i}-\frac{1}{18}F_{\omega_i}C_qM^4_{\omega_i}+A_{\phi\omega,i}^{\eta}M^4_{\omega_i} \nonumber \\
 &&+\frac{\sqrt{2}}{9}C_sF_{\phi_i}M^4_{\phi_i}  
  -\frac{A_i^{\eta}F_{\phi_i}}{6}\big(\frac{5}{3}C_q-\frac{\sqrt{2}}{3}C_s\big)M^4_{\phi_i}
 \nonumber \\
 &&+ A^{\eta}_{\phi\omega,i}M^4_{\phi_i} + 2A^{\eta}_{\phi\omega,i}M^2_{\phi_i}M^2_{\omega_i}
 \bigg) + O\bigg(\frac{1}{t^3}\bigg)\, ,
\end{eqnarray}

\begin{eqnarray}
 &&\kern-30ptF_{\gamma^*\gamma\eta^{'}}(t,0) = \sum^n_{i=1}\frac{4\sqrt{2}h_{V_i}f_{V_i}}{3f_{\pi}} \frac{1}{t} \nonumber \\ 
 &&\bigg(3C_q^{'}M^2_{\rho_i}+\frac{1}{3}F_{\omega_i}C_q^{'}M^2_{\omega_i}  
  + \frac{2\sqrt{2}}{3}C_s^{'}F_{\phi_i}M^2_{\phi_i} \nonumber\\ 
 && + \big(\frac{5}{3}C_q^{'}+\frac{\sqrt{2}}{3}C_s^{'}\big)A_i^{\eta^{'}}F_{\phi_i}M^2_{\phi_i} \bigg) + O\bigg(\frac{1}{t^2}\bigg)\, ,
\end{eqnarray}

\begin{eqnarray}
 &&\kern-20ptF_{\gamma^*\gamma^*\eta^{'}}(t,t) = \sum^n_{i=1}\frac{-8\sqrt{2}h_{V_i}f_{V_i}}{f_{\pi}} \frac{1}{t^2}\nonumber \\ 
 &&\bigg( \frac{1}{2}C_q^{'}M^4_{\rho_i}+\frac{1}{18}F_{\omega_i}C_q^{'}M^4_{\omega_i}
 +\frac{\sqrt{2}}{9}C_s^{'}F_{\phi_i}M^4_{\phi_i}\nonumber \\ 
 && +\frac{A_i^{\eta^{'}}F_{\phi_i}}{6}\big(\frac{5}{3}C_q^{'}+\frac{\sqrt{2}}{3}C_s^{'}\big)M^4_{\phi_i}
 \bigg) + O\bigg(\frac{1}{t^3}\bigg)\, .
\end{eqnarray}

The models are compared often by comparing the slopes of the form factors
 at the origin, which we denote as $a_{P}$.
For the pseudoscalar transition form factors they are defined as: 
\begin{eqnarray}
a_{P} \equiv \frac{1}{F_{\gamma^* \gamma^* P}(0,0)} \frac{d F_{\gamma^* \gamma^* P}(t,0)}{d x} \bigg|_{t=0}
\label{slope}
\end{eqnarray}
where $x\equiv\frac{t}{m^2_P}$.
The model predictions for the model developed in this
 paper read:

\begin{eqnarray}
a_{\pi^0} &=& \frac{16\sqrt{2}\pi^2 m_{\pi^0}^2}{N_c}\sum^{3}_{i=1} h_{V_i} f_{V_i} \nonumber \\
&&\bigg( \frac{1}{M^2_{\rho_i}} + F_{\omega_i}H_{\omega_i}\frac{1}{M^2_{\omega_i}} + A^{\pi^0}_i F_{\phi_i}\frac{1}{M^2_{\phi_i}} \bigg)
\end{eqnarray}
\begin{eqnarray}
a_{\eta} &=& \frac{16\sqrt{2}\pi^2 m_{\eta}^2}{N_c (\frac{5}{3} C_q - \frac{\sqrt{2}}{3}C_s)} \sum^{3}_{i=1} h_{V_i} f_{V_i} \nonumber \\
&&\bigg( 3 C_q\frac{1}{M^2_{\rho_i}} 
+ \frac{1}{3}F_{\omega_i}C_q\frac{1}{M^2_{\omega_i}} 
- \frac{2\sqrt{2}}{3} C_s F_{\phi_i}\frac{1}{M^2_{\phi_i}} \nonumber \\
&&+ (\frac{5}{3}C_q-\frac{\sqrt{2}}{3}C_s)A^{\eta}_i F_{\phi_i}\frac{1}{M^2_{\phi_i}} \bigg)
\end{eqnarray}
\begin{eqnarray}
a_{\eta^{'}} &=& \frac{16\sqrt{2}\pi^2 m_{\eta^{`}}^2}{N_c (\frac{5}{3} C^{'}_q + \frac{\sqrt{2}}{3}C^{'}_s)} \sum^{3}_{i=1} h_{V_i} f_{V_i} \nonumber \\
&&\bigg( 3 C^{'}_q\frac{1}{M^2_{\rho_i}} 
+ \frac{1}{3}F_{\omega_i}C^{'}_q\frac{1}{M^2_{\omega_i}} 
+ \frac{2\sqrt{2}}{3} C^{'}_s F_{\phi_i}\frac{1}{M^2_{\phi_i}} \nonumber \\
&&+ (\frac{5}{3}C^{'}_q+\frac{\sqrt{2}}{3}C^{'}_s)A^{\eta^{'}}_i F_{\phi_i}\frac{1}{M^2_{\phi_i}} \bigg)
\end{eqnarray}

The numerical comparison between predictions within 
  different models and direct extractions from recent experiments is made in Table \ref{tab6}.
 The obtained results are in fair agreement with both. 

\begin{table}[h]
\begin{center}
\vskip0.3cm
\resizebox{0.48\textwidth}{!}{\begin{tabular}{|c|c|c|c|}
\hline
 Model & $a_{\pi^0}$ & $a_{\eta}$ & $a_{\eta^{'}}$ \\
\hline
 fit 1     & 0.0298(3)    &     0.542(4)   &  1.357(9)     \\
 fit 2     &     0.0310(7)      &  0.536(11) &   1.39(3)  \\
 \cite{Czyz:2012nq} &   0.02870(9) &    0.521(2)      &   1.323(4)    \\
\cite{Ametller:1991jv} & 0.0324& 0.506 & 1.470\\
 \cite{Hanhart:2013vba}            &   -    &  0.62+0.06-0.03    &     -   \\
    \cite{Escribano:2013kba}       &   - & $0.60(6)_{st}(3)_{sy}$   &             $1.30(15)_{st}(7)_{sy}$      \\
     \cite{Escribano:2015nra}   &       - &   $0.576(11)_{st}(4)_{sy}$        &             -      \\
 \cite{Escribano:2015yup}      &    -      &    -      &  1.31(4)       \\
CELLO  \cite{Behrend:1990sr}     &   0.0326(26)    &   0.428(63)       &    1.46(16)     \\
SINDRUM-I \cite{MeijerDrees:1992qb}     &   $0.026(24)_{st} (48)_{sy}$   &    -     &     -    \\
\cite{Farzanpay:1992pz} & $0.025(14)_{st} (26)_{sy}$ &- & -\\
 Mami  \cite{Berghauser:2011zz}     &   -   &   $0.576(105)_{st} (39)_{sy}$      &    -     \\
 NA60  \cite{Usai:2011zza}     &   -   &    $0.585(18)_{st} (13)_{sy} $    &     -    \\
  NA62  \cite{TheNA62:2016fhr}     &   $0.0368(51)_{st}(25)_{sy}$     &    -      &  -       \\
\hline
\end{tabular}}
\caption{The slope parameter $a_P$ (Eq.(\ref{slope})) compared to other
  model predictions and experimental data.}
\label{tab6}
\end{center}
\end{table}

\section{Pseudoscalar contributions to $a_\mu$}\label{amu}
 \begin{table}[ht]
\begin{center}
\vskip0.3cm
\begin{tabular}{|c|c|c|c|c|}
\hline
 Model & $a_{\mu}^{\pi^0}$ & $a_{\mu}^{\eta}$ & $a_{\mu}^{\eta^{'}}$ & $a_{\mu}^{P}$ \\
\hline
  fit 1                                 & $58.80\pm0.27$         & $13.56\pm0.10$         & $12.97\pm0.09$         & $85.32\pm 0.30$ \\
  fit 2                                 & $56.96\pm0.94$         & $13.35\pm0.45$         & $12.55\pm0.48$         & $82.85\pm 1.15$ \\
fit 3                                 & $59.07\pm0.17$         & $13.52\pm0.09$         & $12.96\pm0.09$         & $85.55\pm0.22$ \\
 fit 4                                 & $57.79\pm0.90$         & $13.31\pm0.19$         & $12.31\pm0.21$         & $83.41\pm0.94$ \\
\cite{Hayakawa:1997rq}                 & $57.4\pm6.0$                 & $13.4\pm1.6$          & $11.9\pm1.4$                 & $82.7\pm6.4$ \\
\cite{Knecht:2001qf}                  & $58\pm10$                  & $13\pm1$                 & $12\pm1$                 & $83\pm12$\\
\cite{Bijnens:2001cq}                  & -                        & -                          & -                         & $85\pm13$\\
\cite{Melnikov:2003xd}                  & $76.5\pm6.5$                & $18\pm1.4$                & $18\pm1.5$                & $114\pm10$\\
\cite{Dorokhov:2008pw}                  & $62.7-66.8$                 & -                         & -                         & - \\
\cite{Nyffeler:2009tw,Jegerlehner:2009ry}                 & $72\pm12$                 & $14.5\pm4.8$                 & $12.5\pm4.2$                 & $99\pm16$ \\
\cite{Kampf:2011ty}                  & $68.8\pm 1.2$                 & -                        & -                         & - \\
\cite{Roig:2014uja}                  & $66.6 \pm 2.1$                 &    $20.4\pm 4.4$                    &       $17.7\pm2.3$                  & $104.7\pm 5.4$  \\
\cite{Gerardin:2016cqj}                  & $65.0\pm8.3$                 & -                        & -                         & - \\
 \hline
\end{tabular}
\caption{Pseudoscalar-exchange contribution to the $a_{\mu}^{HLBL,PS} \times 10^{11}$ ($PS=\pi^0, \eta, \eta^{'}$). }
\label{tab5}
\end{center}
\end{table}

  Within the model described in the previous sections we calculate the contributions from
 the pseudoscalar mesons $\pi^0, \eta$ and $\eta'$ to the muon anomalous magnetic moment
  $a_\mu$.
 The formula Eq.(155) of \cite{Jegerlehner:2009ry} was used with the form factors developed
  in this paper Eqs.(\ref{ffpi0}-\ref{ffetap}). The variables spanned from zero to infinity 
 were mapped on the intervals $(0,1)$ and the integrals were performed using the Monte Carlo method.
 For a cross check of the numerical method and the implementation we have recovered values
  from Table 7 of \cite{Jegerlehner:2009ry} using the model(s) presented there.
  The results are presented in Table  \ref{tab5} for both fits and compared with previous
  calculations.
  For the error evaluation we have used the covariance matrix calculated by Minuit from CERNLIB.
   The derivatives of the $a_\mu$ in respect to the fitting parameters were calculated 
   numerically,  using the Monte Carlo method to obtain the necessary integrals.
  The error of the sum of all the contributions from pseudoscalars was 
  calculated separately as an error on the function being the sum of the
  free contributions. 
   As one can observe the obtained results are consistent with most of other models.
   The biggest differences,  not contained in the error bars, are observed with 
   calculations presented in \cite{Melnikov:2003xd,Kampf:2011ty,Roig:2014uja}. 
  The much smaller errors of our calculations, as compared to other results,
    are only
   parametric and do not cover the model dependence. Yet, it has to be stressed that the 
  model is able to describe well all the existent data on the form factors both
   in the space-like and time-like regions. To cover the  model dependence within
   the class of models we consider here we added two values of $a_\mu$ (fit 4 and fit 5).
   In the models 4 and 5 we have excluded
   from the fit the cross sections of the reactions  $e^+e^-\to \eta \gamma$ and $e^+e^-\to\eta' \gamma$
   measured by BaBar \cite{Aubert:2006cy} at very high energy compared to other
   data points. The fits were performed 
   with parameters $A_3^P$ set to zero and with fixed or fitted mixing parameters
   similarly to fits 1 and 2. The $e^+e^-\to\eta' \gamma$ cross section calculated at the BaBar energy point
   is off the measured value by about 5 standard deviations.
    Also the predicted $e^+e^-\to\pi^0 \gamma$ cross section at $s=112\ {\rm GeV}^2$
   is different for both fits. However, as expected from the analysis in \cite{Nyffeler:2016gnb},
    the values of the pseudoscalar form factors at large invariant masses are much less important
    than the behaviour in the range up to about 1 GeV for the calculation of $a_\mu$.
    Thus the very close results for  $a_\mu$ coming from all the fits are not surprising.
    The range of the predicted values of $a_\mu$ within the class of models we examined
    is thus $ 79.4\times 10^{11} <a_\mu^P< 86.23\times 10^{-11}$, if we take conservatively 
    $3\sigma$ errors, and
     the predicted value of $a_\mu$ is $(82.8\pm 3.4)\times 10^{-11}$.
\

\section{The implementation of the model in Ekhara and Phokhara generators}\label{phek}

\begin{figure}
\begin{center}
\hskip-1cm\includegraphics[width=9.cm]{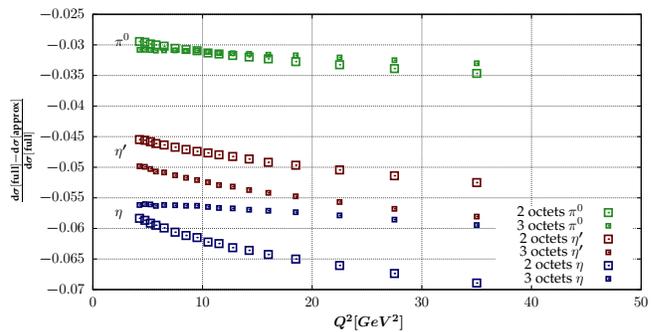}
\caption{The relative difference of differential cross sections calculated
  with $F_{\gamma^*\gamma^* P}(-Q^2,q_1^2)$ (full)
  and $F_{\gamma^*\gamma^* P}(-Q^2,0)$ (approx). See text for details.
}
\label{twovirt}
\end{center}
\end{figure}

 The new transition pseudoscalar form factors were implemented in the
 event generator EKHARA \cite{Czyz:2010sp,Czyz:2006dm}. As one can see
 from Figure \ref{plot1} the difference of the form factors
 from this paper as compared to the old model \cite{Czyz:2012nq},
 for the configuration, where one of the invariant is equal to zero,
  is not big. Yet, the experiments never have the second invariant mass
 equal to zero and the events are collected with a cut resulting from
  the cuts on the observed particles. The influence of this effect 
  on the experimental side is a part of the systematic error. On the theory
  side it is model dependent, with the part which is different from zero
  only when both photon virtualities are different from zero never
   tested directly by any experiment in the space-like region.
   The difference of the predictions of the
   influence of the second virtuality between the old and the new model 
  is shown in Figure \ref{twovirt}. We plot there the relative difference 
  of the differential cross sections calculated with the complete form factors
  (full) and the case where one of the invariants was set to zero (approx) as
   a function of the second invariant $Q^2=-(q-p)^2$.
   $q$ is the four-vector of the final positron and 
  $p$ is the  four-vector of the initial positron.
   We limit the invariant mass squared of the first
  virtual photon  ($Q_1^2=-(q'-p')^2$, where $q'$ is the four-vector of the final electron and 
  $p'$ is the  four-vector of the initial electron) to $Q_1^2\leq 0.18$~GeV for $\pi^0$ and to $Q_1^2\leq 0.38$
   for $\eta$ and $\eta'$. As one can see the corrections coming from
  the second invariant are by no means negligible, and their size
  exhibits the model dependence. 
  In the plot the form factors of the 'fit 2' were used. For the 'fit 1'
  they look similar.

 Having the model of the pseudoscalar transition form factors 
  valid also in the time-like region we are able to 
 simulate  the cross sections of the reactions $e^+e^- \to P\gamma$.
 This is done within the Phokhara Monte Carlo generator \cite{Rodrigo:2001kf}
  framework. It is an upgrade of the version 9.2 \cite{Czyz:2016xvc}
  and will be available from the web page
 {\it (http://ific.uv.es/$\sim$rodrigo/phokhara/)} 
   as release 9.3. Both options with the fit 1 and the fit 2 parameters
  are implemented.
   The next to
 leading order initial state radiative corrections  
  were included basing on the approach described in \cite{Czyz:2013xga}.
 The virtual and the soft initial state corrections are universal and
  are exactly the same as in \cite{Czyz:2013xga}, thus we do not repeat 
  here the formulae.
The matrix element describing the reaction $e^+ e^- \to \pi^0 \gamma \gamma $ 
 was written as a product of leptonic and hadronic current:
\begin{eqnarray}
&&\mathcal{M}[e^+(p_1) e^-(p_2)\to \pi^0(q_1) \gamma(k_1) \gamma(k_2) ] 
  \nonumber \\
  && = L^{\nu}(k_1)H_{\nu}(k_2) + (k_1 \leftrightarrow k_2),
\end{eqnarray}
where
\begin{equation}
H_{\nu}(k_2)=e^2 \epsilon_{\mu\nu\alpha\beta} q_{1}^{\mu} k_{2}^{\alpha} 
  \epsilon_2^{\beta}
F_{\gamma^*\gamma^* P}\big((q_1+k_2)^2, 0\big)
\end{equation}
and
\begin{eqnarray}
L^{\nu}(k_1)&=& \frac{ie^2}{2p_2\cdot k_1} \bar{v}(p_1)\gamma^{\nu}\Big(2\epsilon_1p_1-\ta{k_1}\ta{\epsilon_1}\Big)u(p_2) \nonumber\\
&& + \frac{ie^2}{2p_1\cdot k_1} \bar{v}(p_1)\Big(\ta{\epsilon_1}\ta{k_1}-2\epsilon_1 p_1\Big)\gamma^{\nu}u(p_2)\nonumber,\\
\end{eqnarray}
  with $\epsilon_i, i=1,2$ being a polarization vector of the photon with
  the four momentum  $k_{i}$.

\begin{figure}
\begin{center}
\hskip-1cm\includegraphics[width=9.cm]{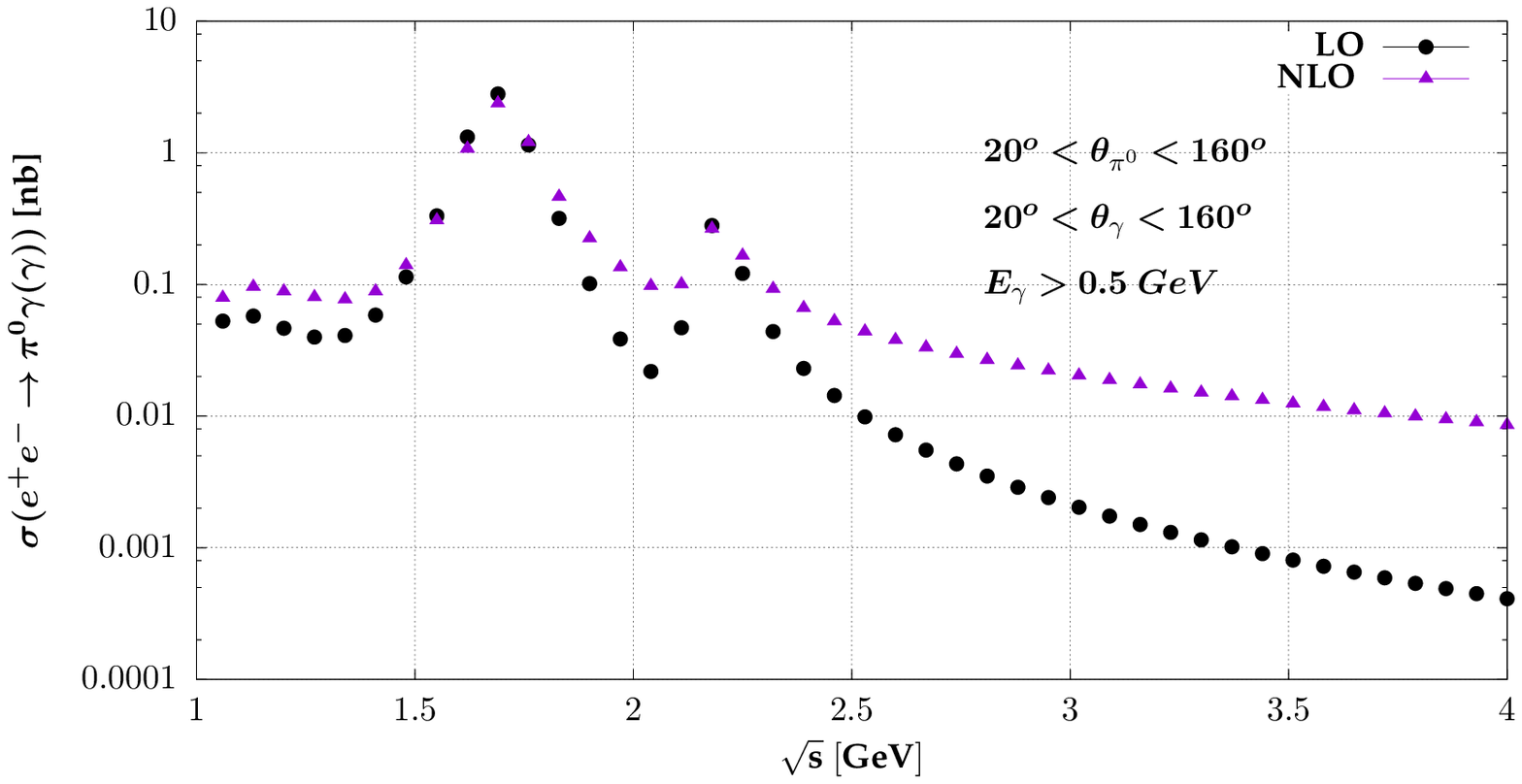}
\phantom{}\hskip-1cm\includegraphics[width=9.cm]{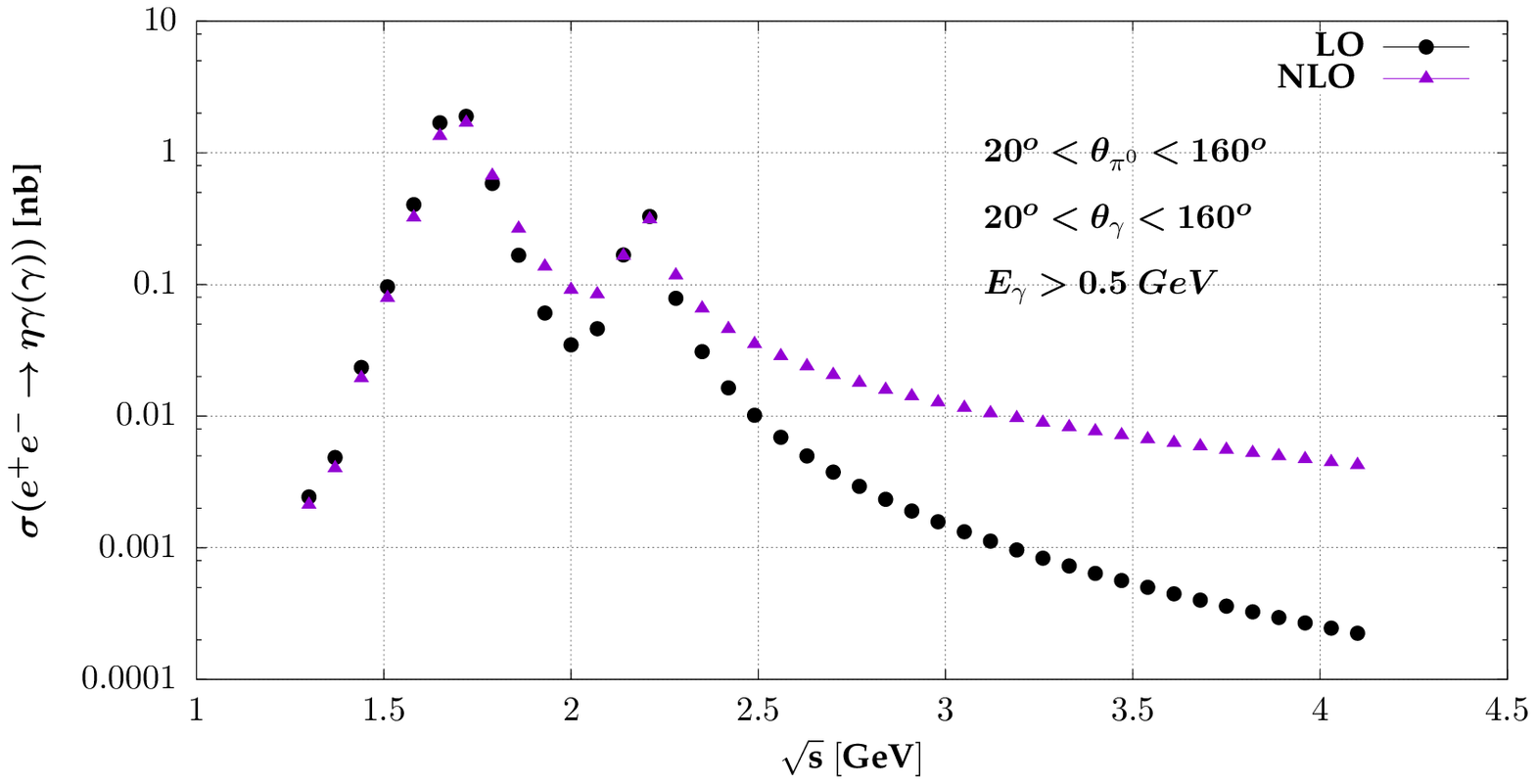}
\phantom{}\hskip-1cm\includegraphics[width=9.cm]{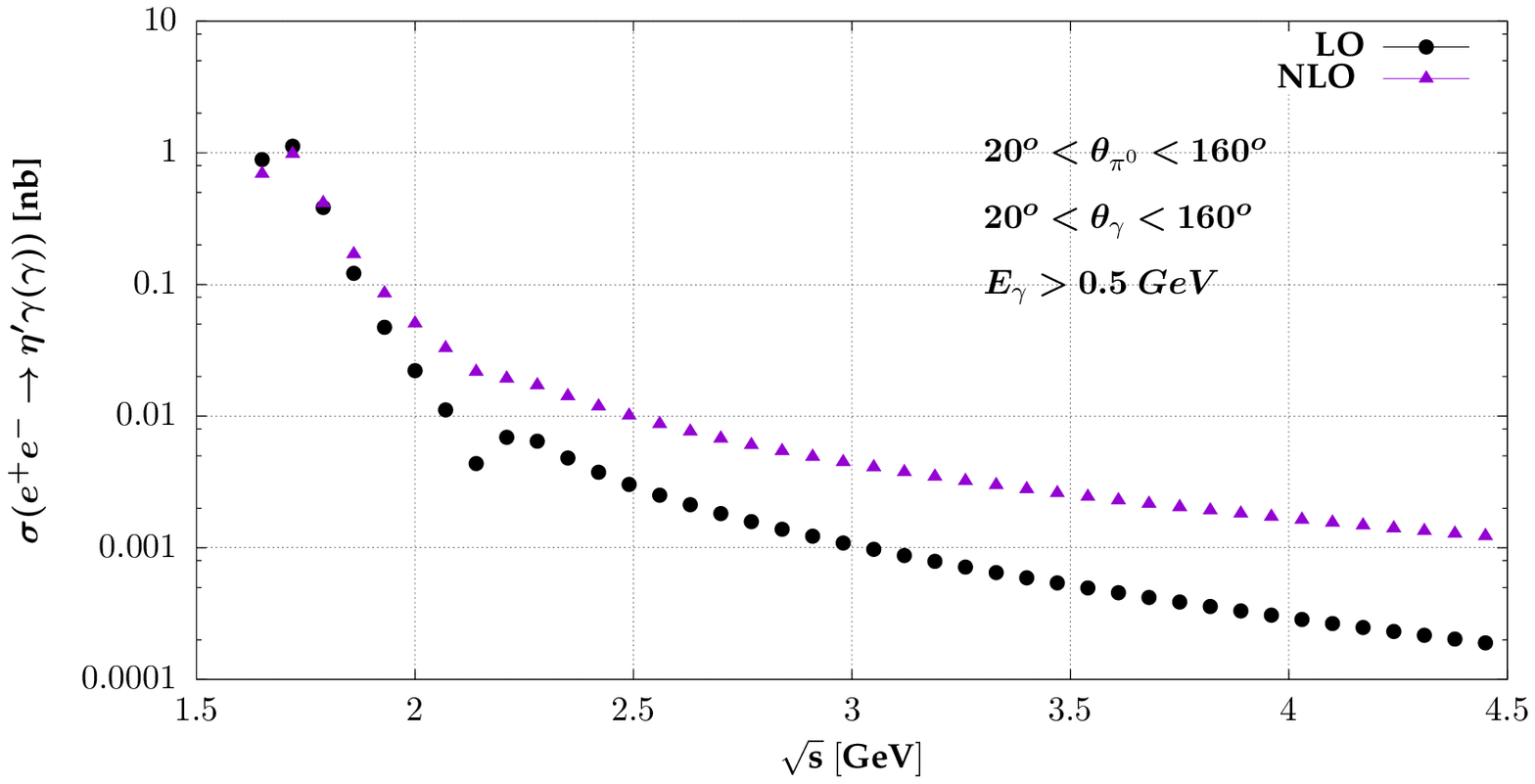}
\caption{Comparison between LO and NLO cross sections. See text for details.
}
\label{pigam}
\end{center}
\end{figure}

   The effect of radiative corrections is shown
   in Figure \ref{pigam}.
  The plots were obtained using fit 2 parameters
    accepting the events with the pseudoscalar
   particle  and
  one of the photons with an energy bigger than 0.5 GeV being observed within
  the angular range between 20 and 160 degrees.
   The radiative corrections are big due to the fact that 
    the pseudoscalar transition form factor is falling fast at high values
   of the virtual photon mass. At LO (leading order) the form factor is
  calculated at $s$, while in the two photon amplitude it is calculated
  at much smaller invariants $Q_1^2=(q_1+k_1)^2$, or $Q_2^2=(q_1+k_2)^2$ resulting
   from the hard photon emission.

\section{Conclusions}\label{conc}

We model the Lagrangians
 $\mathcal{L}_{\gamma \gamma \mathcal{P}}$,
    $\mathcal{L}_{\gamma V}$, $\mathcal{L}_{ V\gamma P}$ and  $\mathcal{L}_{VVP}$
   within the resonance
  chiral symmetric theory with the SU(3) breaking. Two model versions with
    22(17) couplings of the model are fitted 
  to 536 experimental data points resulting in $\chi^2=415(454)$. 
   Within the developed models we predict the light-by-light
  contributions to the muon anomalous magnetic moment $a_{\mu}^P = (82.8 \pm 3.4)\times 10^{-11}$.
  The error covers also the model dependence within the class of models considered in this paper.
  The model was implemented into the Monte Carlo event generator Ekhara to simulate the
  reactions $e^+e^-\to e^+e^- P$, ($P=\pi^0,\ \eta,\ \eta'$) and into the Monte Carlo event generator
  Phokhara to simulate the reactions $e^+e^-\to P\gamma(\gamma)$ at the next-to-leading order.

\bibliography{biblio}

\end{document}